\documentclass[12pt]{article}
\usepackage{mathrsfs}
\usepackage{amsmath}
\usepackage{mathrsfs}
\usepackage{amssymb}
\usepackage{color}
\usepackage{psfrag}
\usepackage{amsmath}
\usepackage{amssymb}
\usepackage{graphicx}
\usepackage{subfigure}
\usepackage{cite}

\newcommand{\bea}{\begin{eqnarray}}
\newcommand{\eea}{\end{eqnarray}}
\newcommand{\be}{\begin{equation}}
\newcommand{\ee}{\end{equation}}
\newcommand{\vs}[1]{\vspace{#1 mm}}

\renewcommand{\b}{\beta}

\newcommand{\dsl}{\pa \kern-0.5em /}

\newcommand{\half}{\frac{1}{2}}
\newcommand{\pa}{\partial}

\newcommand{\nn}{\nonumber\\}

\newcommand{\ba}{\begin{array}}
\newcommand{\ea}{\end{array}}
\newcommand{\bit}{\begin{itemize}}
\newcommand{\eit}{\end{itemize}}
\newcommand{\tp}{\triangle _+}
\newcommand{\tm}{\triangle _-}
\begin{document}
\topmargin 0mm
\oddsidemargin 0mm

\begin{flushright}

USTC-ICTS-16-03\\

\end{flushright}

\vspace{2mm}

\begin{center}

{\Large \bf Phase structures of 4D stringy charged black holes\\
 in canonical ensemble}

\vs{10}

{\large Qiang Jia,   J. X. Lu and Xiao-Jun Tan}

\vspace{4mm}

{\em

 Interdisciplinary Center for Theoretical Study\\
 University of Science and Technology of China, Hefei, Anhui 230026, China\\
}

\end{center}

\vs{10}

\begin{abstract}
We study the thermodynamics and phase structures of the asymptotically flat dilatonic black holes in 4 dimensions, placed in a cavity {\it a la} York, in string theory for an arbitrary dilaton coupling. We consider these charged black systems in canonical ensemble for which the temperature at the wall of and the charge inside the cavity are fixed. We find that the dilaton coupling plays the key role in the underlying phase structures. The connection of these black holes to higher dimensional brane systems via diagonal (double) and/or direct dimensional reductions indicates that the phase structures of the former may exhaust all possible ones of the latter, which are  more difficult to study, under conditions of similar settings.  Our study also shows that a diagonal (double) dimensional reduction preserves the underlying phase structure while a direct dimensional reduction has the potential to change it. 
 \end{abstract}

\newpage



\section{Introduction}

It is known that an asymptotically flat black hole is thermodynamically unstable due to its Hawking radiation. There are in general two ways to resolve the instability issue. One is to add a negative cosmological constant to the system such that the resulting black hole is in AdS space\cite{Hawking:1982dh}. The other is to place the asymptotically flat spherical black hole in a spherical cavity outside its horizon with the temperature of the cavity wall fixed {\it a la} York\cite{York:1986it}.  Our interest in this paper is for the latter case. Further when the charge inside the cavity is fixed, we define an canonical ensemble while the fixed potential\cite{Braden:1990hw} on the wall of the cavity defines the so-called grand canonical ensemble. 

We will study the thermodynamics and phase structures of various 4-dimensional asymptotically flat black holes in string theory given in \cite{Garfinkle:1990qj} with an arbitrary dilaton coupling, following the aforementioned cavity approach in canonical ensemble. We ask the following questions: if the temperature at the wall of the cavity and the charge inside it are fixed, what are possible thermally stable phase structures allowed?   How do the underlying phase structures depend on the dilaton coupling?  What are the possible connections of these phase structures to those of branes in 10 dimensional string theory and 11 dimensional M-theory?  Does the dimensional reduction (either direct or diagonal (double) dimensional one) change or preserve the underlying phase structure?   We will explore all in this paper. 

We first examine the chargeless case. We find that the dilaton coupling has no effect on the phase structure for this case and the underlying phase structure remains precisely the same as that of non-dilatonic case studied before in \cite{Lundgren:2006kt} and that of higher dimensional chargeless branes \cite{Lu:2010xt}. In other words, there exists a minimal temperature below which the only thermally stable phase inside the cavity is the so-called hot empty space. Above this minimal temperature, we have in general a small and a large black holes but only the large one is locally stable. Further there exists a transition temperature above the minimal one at which the large locally stable black hole has zero Helmholtz free energy, just as the hot empty space at the same temperature. As such the two phases can coexist and the phase transition between the two is just a first order one since the transition involves an entropy change. For temperature lower than this but still higher than the minimal one, the locally stable black hole has a positive free energy, therefore it is globally unstable and will make a so-called Hawking-Page transition to the hot empty space at the same temperature. Only for the temperature higher than the transition temperature, the locally stable black hole has a negative Helmholtz free energy and becomes a globally stable phase.       

We then examine the charged case. We find that when the dilaton coupling is  greater than one, the underlying phase structure resembles the one of the chargeless case except that the hot empty space is now replaced by the corresponding extremal black hole with the preset temperature.  In particular, this type of phase structure is also essentially the same as that of charged 6-brane case studied in \cite{Lu:2010xt}. When the dilaton coupling is equal to unity,  this situation resembles that of charged 5-brane case studied also in \cite{Lu:2010xt} for which there exists a critical charge but there is no critical second-order phase transition. There are three sub-cases but  the phase structure for each of them resembles that of chargeless case, without  the presence of the van-der Waals-Maxwell gas-liquid type one, again with the replacement of hot empty space by the corresponding extremal black hole.  However, when the dilaton coupling is less than one, we find that there exists a critical charge\footnote{Without loss of generality, we assume $q \ge 0$ from now on.}  $q_c$ above which we have a globally stable black hole at every preset temperature. Below this critical charge, we find maximal and minimal temperatures between which there exist three black hole phases. The largest and smallest black holes are locally stable as their free energies are local minima while the intermediate size black hole is unstable as its free energy is a local maximum. Further the free energies of the two locally stable black holes are different and there exists a transition temperature, for any given charge $q < q_c$, at which their free energies are the same. At this temperature, the small and the large black holes can coexist and transit freely from one phase to the other. For a temperature lower (higher) than the transition temperature, the small (large) black hole has a lower free energy and is globally stable. This transition is a first order phase transition since it involves an entropy change. When $q = q_c$, the two stable black holes coincide therefore the phase transition ends up at a second order phase transition point. This kind of phase structure is nothing but the familiar van der Waals-Maxwell gas-liquid type which exists for the usual Reissner-Nordstr$\ddot {\rm o}$m (charged) black hole, i.e, the zero dilaton coupling case studied in \cite{Lundgren:2006kt}, charged black p-branes in \cite{Lu:2010xt} for $0 \le p \le 4$, the (D1, D5) (or (F, NS5)) in \cite{Lu:2012rm} and the (D0, D6) system in \cite{Lu:2013nt, Lu:2014dra} in 10-dimensional string theories, and the black M-branes in 11-dimensional M-theory \cite{Lu:2010xt} .  Given this, one may suspect that the phase structures of 4-dimensional charged black holes with different dilaton couplings follow just those of the aforementioned charged black p-banes since after all they appear to be related via dimensional reductions. This turns out to be true.  The higher dimensional systems with three overall transverse dimensions (the same as the 4-dimensional charged black holes) are connected to the corresponding 4-dimensional charged black holes via purely diagonal (double) dimensional reductions and as such the former shares characteristically the same phase structure as the latter.  Previous studies indicate that different kinds of such higher dimensional systems in a given bulk dimension $D \ge 4$  can capture all possible phase structures in that dimension D under conditions of similar settings. Combining these two, we may reach  that the phase structures of  4-dimensional charged black holes with different dilaton couplings exhaust all possible ones of higher dimensional charged black brane systems\footnote{By saying this, we exclude the so-called bubble phase given in \cite{Lu:2011da}, which can occur for charged black p-brane systems with $p \ge 1$.}, simple or complicated/known or unknown, under conditions of similar settings. This gives an advantage and usefulness of the  present study.  In general, we also find that the phase structure is preserved by a diagonal (double) dimensional reduction but this is not true if a direct dimensional reduction is involved.  

For 4D stringy black holes with dilaton coupling less than unity, we calculate the corresponding critical charge $q_c$, the black hole size $x_c$ and the transition inverse 
temperature  $b_c$ for different fixed dilaton couplings. 

This paper is organized as follows. In section 2, we briefly review the 4-dimensional asymptotically flat black holes in string theory given in \cite{Garfinkle:1990qj} and derive the Euclidean action which we employ to study the thermodynamics and phase structures. In section 3, we analyze the general thermal stability of black holes in a cavity in canonical ensemble. Section 4 discusses the phase structures of the chargeless case and charged case, respectively.  The relation of 4D stringy charged black holes to higher dimensional brane systems including the 10 and 11 dimensional ones in phase structures is discussed in detail  in section 5.  We conclude this paper in section 6. 

\section{4D stringy black holes and their free energies}

The 4-dimensional bosonic part of the low-energy effective action\footnote{In the action, we have set the 4D Newton constant $G = 1$.} of string theory is \cite{Garfinkle:1990qj}
\be\label{ea}
S= \frac{1}{16 \pi} \int d^4 x \sqrt{-g} \left(\mathcal{R}-2(\nabla \phi)^2 -e^{2 a\phi}F^2\right),
\ee
where $\phi$ is the dilaton and $a$ is the dilaton coupling.  $F$ is a $U(1)$ gauge field strength. The respective equation of motion for each field can be obtained from the above action.  The general black hole solution given in \cite{Garfinkle:1990qj} is
\bea \label{bhsolution}
	&& ds^2= - \lambda ^2 dt^2+\frac{dr^2}{\lambda^2}+R^2d\Omega^2_2, \nn
	&&A =  \frac{e^{-a\phi_0}}{(1+a^2)^{\half}} \left[ \left(\frac{r_-}{r_+}\right)^{\half} - \frac{(r_- r_+)^\half} {r} \right] dt, \nn
	&&F =  \frac{e^{-a\phi_0}}{(1+a^2)^{\half}}\frac{(r_{+}r_{-})^{\half}}{r^2}dr\wedge dt, \nn
	&&e^{- 2 a (\phi - \phi_0)} = \left(1 - \frac{r_-}{r}\right)^{\frac{2 a^2}{1 + a^2}},	
\eea 
where
\bea\label{bhone}
&&\lambda^2=\left(1-\frac{r_+}{r}\right)\left(1-\frac{r_-}{r}\right)^{\frac{1-a^2}{1+a^2}}	\nn
&&R=r\left(1-\frac{r_-}{r}\right)^{\frac{a^2}{1+a^2}}.
\eea
In the above, $r_+$ and $r_-$ are coordinate radii of the outer and inner horizons, respectively, and $\phi_0$ is the asymptotic value of dilaton. The metric represents a black hole only for $r_+ > r_-$.  For latter convenience, we have expressed the above solution as an electric one which is related to the original magnetic one given in \cite{Garfinkle:1990qj} for the gauge field via  a hodge duality.  The 1-form gauge potential $A$ defined above is shifted by a constant, following \cite{York:1986it}, such that it vanishes on the the outer horizon and as such it is well-defined in the local inertial frame.

For the purpose of studying the thermodynamics\cite{York:1986it}, we consider the corresponding solutions given in Euclidean signature via $t \to - i \tau$ as,


\bea \label{emetric}
	&& ds^2=\lambda ^2 d\tau^2+\frac{dr^2}{\lambda^2}+R^2d\Omega^2_2 \nn
	&&A = - i \frac{e^{-a\phi_0}}{(1+a^2)^{\half}} \left[ \left(\frac{r_-}{r_+}\right)^{\half} - \frac{(r_- r_+)^\half} {r}\right] d\tau \nn
	&&F = - i \frac{e^{-a\phi_0}}{(1+a^2)^{\half}}\frac{(r_{+}r_{-})^{\half}}{r^2}dr\wedge d\tau 	
\eea
where $\phi$, $\lambda^2$ and $R$ remain the same as those given in (\ref{bhsolution}) and (\ref{bhone}) .   For the metric free of a conical singularity at $r=r_+$, we need to have the Euclidean time $\tau$ with  a periodicity,
\be\label{iT}
\beta ^*=4\pi r_+ \left(1-\left(\frac{r_-}{r_+}\right)\right)^{-\frac{1-a^2}{1+a^2}},
\ee
which is the inverse temperature at $r=\infty$. The local inverse temperature $\beta $ at a given radius $r$ can be obtained from $\beta^*$ as,
\begin{align}\label{lT}
\beta(r) &=\lambda \beta^* \nonumber \\
&=\left(1-\frac{r_+}{r}\right)^{\frac{1}{2}}\left(1-\frac{r_-}{r}\right)^{\frac{1-a^2}{2(1+a^2)}}\beta ^*  ,
\end{align}
where $\lambda$ is the redshift factor. We place this black hole inside a cavity located at $r = r_B > r_+$ with its corresponding physical radius 
\be
R_B = \left(1-\frac{r_-}{r_B}\right)^{\frac{a^2}{1+a^2}}r_B,
\ee
where we have used the expression for $R$ given in (\ref{bhone}).
It is this physical radius $R_B$ that we fix in canonical ensemble.  Similarly, we define
\bea
&&R_{\pm}=\left(1-\frac{r_-}{r_B}\right)^{\frac{a^2}{1+a^2}}r_{\pm} \nn
&&\triangle_{\pm}=1-\frac{r_{\pm}}{r_B}=1-\frac{R_{\pm}}{R_B}
\eea
and they are the parameters which we will use from now on. In canonical ensemble we fix the local temperature at the wall of cavity
\be	\label{temperature}
\b \equiv \b(R_B)=\tp^{\half}\tm^{\frac{1-a^2}{2(1+a^2)}}\b ^*
\ee
and the charge inside the cavity  as\footnote{Without loss of generality, we assume from now on $Q \ge 0$.},
\begin{align}\label{charge}
Q&= - \frac{i}{4\pi}\int e^{2 a \phi}*F=e^{a \phi_0}\left(\frac{r_+r_-}{1+a^2}\right)^{\frac{1}{2}} \nn
&=e^{a \phi_B}\left(\frac{R_+ R_-}{1+a^2}\right)^{\frac{1}{2}}
\end{align}
where in the second line we have used the dilaton expression given in (\ref{bhsolution}) to convert the asymptotic value of dialton $\phi_0$ to its value $\phi_B$ at $R_B$ which is fixed for the canonical ensemble considered.  By this, the coordinate parameters $r_\pm$ are changed to the corresponding physical ones $R_\pm$. 

For the purpose of obtaining the leading order contribution to the Helmholtz free energy for the black hole system considered, we follow the standard procedure given in \cite{York:1986it,Gibbons:1976ue,Lu:2010xt} to compute the Euclidean action for the black hole configuration given in (\ref{emetric}). The Euclidean action has three parts 
\be	\label{action}
I_E=I_E(g)+I_E(\phi)+I_E(A) .
\ee
In the above, each part is given, respectively, as
\be
I_E (g) = - \frac{1}{16 \pi} \,\int_M d^4 x \sqrt{g_E} \, R_E  + \frac{1}{8\pi} \,\int_{\partial M} d^3x \sqrt{\gamma}\,(K-K_0), 
\ee
where the second term is the Gibbons-Harking boundary term \cite{Gibbons:1976ue}, the dilaton action
\be 
I_E (\phi) = \frac{1}{8 \pi} \int_M d^4 x \sqrt{g_E}\, (\nabla \phi)^2,
\ee
and finally the action for the electromagnetic field 
\be 
I_E (A)  = \frac{1}{16 \pi} \,\int_M d^4 x \sqrt{g_E}\,  e^{2 a \phi} \,F^2 - \frac{1}{4\pi}\, \int_{\partial M} d^3 x \sqrt{\gamma}\, e^{2a\phi}\,n_{\mu}F^{\mu \nu}A_{\nu},
\ee
where $\pa M$ denotes the boundary of the Euclidean 4-dimensional manifold $M$,  $\gamma$ is the induced metric on the boundary, $n_\mu$ is the normal vector of the boundary
with the normalization $n_\mu n^\mu = 1$, and $K$ is the trace of the extrinsic curvature $K_{\mu \nu}$ defined as 
\be
K=g^{\mu \nu}K_{\mu \nu}=-\frac{1}{2}g^{\mu \nu}(\nabla_{\mu}n_{\nu}+\nabla_{\nu}n_{\mu}) = -  \nabla_{\mu}n^{\mu}.
\ee
In the above, $K_0$ is the trace of the extrinsic curvature of the following flat Euclidean metric at given physical radius $R$ 
\be 
ds^2 = d\tau^2 + R^2 d\Omega^2_2.
\ee

The path integral for a given gravitational system in Euclidean signature corresponds to the thermodynamical partition function  $e^{-\b F}$, where $\b$ is the inverse temperature of the system and $F$ is the Helmholtz free energy in canonical ensemble. The saddle point approximation (zero loop) of the path integral can be obtained as $Z \approx e^{-I_E}$ where $I_E$ is the Euclidean action evaluated for a given configuration (see for example \cite{Brown:1994su}). Therefore we have $F\approx I_E/\b$. Since $\b$ is fixed, so $I_E$ is the quantity which we will use to analyze the phase structure and thermal stability of the black holes in what follows.

We can simplify the evaluation of the Euclidean action \eqref{action} by making use of certain equations of  motion. The equation of motion for metric  is
\be
\mathcal{R}_{\mu \nu} -\frac{1}{2}g_{\mu \nu} \mathcal{R}= 2\nabla_{\mu} \phi \nabla_{\nu} \phi - g_{\mu \nu} (\nabla \phi)^2 +e^{2 a\phi}\left(2F_{\mu \rho}F_{\nu}^{\rho} - \frac{1}{2}g_{\mu \nu} F^2\right).
\ee
From the above, we have
\be	\label{Ricci}
\mathcal{R}=2(\nabla \phi)^2.
\ee
With this, we have
\begin{align}
I_E & = I_E (g)  + I_E (\phi)  + I_E (A) \nn
&= \frac{1}{16 \pi} \,\int_M d^4x \sqrt{g_E} \, e^{2a\phi}\, F^2 + \frac{1}{8\pi}\,\int_{\partial M} d^3x \sqrt{\gamma}\,(K-K_0) - \frac{1}{4\pi}\,\int_{\partial M} d^3 x \sqrt{\gamma}\,e^{2a\phi}\,n_{\mu}F^{\mu \nu}A_{\nu}
\end{align}
where $F_{\mu \nu}$ and $A_{\nu}$ are given in (\ref{emetric}).  The normal vector $n_{\mu}$, the trace of the extrinsic curvatures $K$ and $K_0$ on the boundary can be calculated from the metric  \eqref{emetric}  as
\begin{align}
n^{\mu} & =\triangle_+ ^{\frac{1}{2}} \triangle_- ^{\frac{1-a^2}{2(1+a^2)}}\delta^{\mu}\,_r \nonumber \\
K & =-\nabla _{\mu} n^{\mu} \nonumber \\
  & =-\frac{1}{r_B}\tp ^{\frac{1}{2}} \tm^{\frac{1-a^2}{2(1+a^2)}}\left(\frac{1+3a^2}{2(1+a^2)}\tm ^{-1} +\frac{1}{1+a^2} +\frac{1}{2}\tp^{-1}\right)\nonumber \\
K_0 & =-\frac{2}{r_B} \tm^{-\frac{a^2}{1+a^2}}=-\frac{2}{R_B}.
\end{align}
Now we can evaluate the action to give,
\begin{align}\label{action2}
I_E=&\beta R_B \left(1-\frac{1}{1+a^2}\tp^{\frac{1}{2}}\tm^{\frac{1}{2}}-\frac{a^2}{1+a^2}\tp^{\frac{1}{2}}\tm^{-\frac{1}{2}}\right) \nonumber \\
&- \pi   R_+^2\tm^{\frac{-2a^2}{1+a^2}}\left(1-\frac{R_-}{R_+}\right)^{\frac{2a^2}{1+a^2}}.
\end{align}
Since we have the Helmholtz free energy $F=E-TS$ and $F=I_E/\beta$, this gives $I_E=\beta E-S$ where $E$ is the energy inside the cavity and $S$ is the entropy of the system
\bea \label{entropy}
&&E=  R_B \left(1-\frac{1}{1+a^2}\tp^{\frac{1}{2}}\tm^{\frac{1}{2}}-\frac{a^2}{1+a^2}\tp^{\frac{1}{2}}\tm^{-\frac{1}{2}}\right), \nn
&&S=\pi  R_+^2\tm^{\frac{-2a^2}{1+a^2}}\left(1-\frac{R_-}{R_+}\right)^{\frac{2a^2}{1+a^2}}=\pi  r_+^2\left(1-\frac{r_-}{r_+}\right)^{\frac{2a^2}{1+a^2}}=\frac{ A}{4} .
\eea
Here $A=4\pi r_+^2\left(1-\frac{r_-}{r_+}\right)^{\frac{2a^2}{1+a^2}} $ is the physical area of the outer horizon. Note that taking $R_B 
\to \infty$, the energy becomes the ADM mass given in \cite{Garfinkle:1990qj} as it should be. The above computed entropy is just the one of the black hole without the presence of the cavity,  as this fact was stressed in \cite{York:1986it}. The computed $I_E$ provides the basis for us to analyze the stability of black holes considered in the next section. 

\section{Generalities and stability of the black holes}
As will be seen, the relevant quantities for analyzing the thermal stability of 4D stringy black holes are the following dimensionless ones:
\be
x=\frac{R_+}{R_B} < 1,\,\ q=\frac{Q^*}{R_B},\, \bar b=\frac{\beta}{4\pi R_B},\, \bar I_E=\frac{I_E}{4 \pi R_B^2}
\ee
where $Q^*$ is defined as $Q^*=Q e^{-a \phi_B}$, which is also fixed. Note that 
\be	\label{charge}
x < 1, \qquad q=\frac{Q^*}{R_B} < \left(\frac{1}{1+a^2}\right)^{\half} \frac{R_+}{R_B} =  \left(\frac{1}{1+a^2}\right)^{\half} x,\ee
which further imply 
\be \label{xrange}
\sqrt{1 + a^2} q \le x < 1.
\ee
Note that the lower end limit, i.e. $x \to q\sqrt{1 + a^2}$, corresponds to the extremal limit.
In the canonical ensemble, the quantities $R_B$, $\phi_B$, $Q$ and $\beta$ are fixed. So $q$ and $\bar b$ are fixed, the only variable is $x$, the reduced dimensionless horizon size. 
The reduced action $\bar I_E$ can now be expressed in terms of the above defined dimensionless quantities as,
\begin{align} \label{action3}
\bar I_E =&\bar b \left(1-\frac{a^2 \sqrt{1-x}}{\left(1+a^2\right) \sqrt{1-\frac{\left(1+a^2\right) q^2}{x}}}-\frac{\sqrt{1-\frac{\left(1+a^2\right) q^2}{x}} \sqrt{1-x}}{1+a^2}\right) \nonumber \\
&-\frac{1}{4} \left(1-\frac{\left(1+a^2\right) q^2}{x^2}\right)^{\frac{2 a^2}{1+a^2}} \left(1-\frac{\left(1+a^2\right) q^2}{x}\right)^{-\frac{2 a^2}{1+a^2}} x^2.
\end{align}
From the above, we have
\be \label{partial action}
\frac{\partial \bar I_E}{\partial x}=c(x)\left[\bar b - b_{a, q} (x) \right],
\ee
where $b_{a, q} (x)$ and  $c(x)$ are, respectively,
\bea \label{bf}
b_{a, q} (x) &=& x \sqrt{1-x} \left(1-\frac{(1+a^2)q^2}{x}\right)^{\frac{1-3a^2}{2(1+a^2)}}\left(1-\frac{(1+a^2)q^2}{x^2}\right)^{\frac{a^2-1}{a^2+1}},\nn
c(x) &=& \frac{\left(1+a^2\right) q^4+x^3-q^2 x \left[1+x+a^2 (-1+2 x)\right]}{2\, x^3\, \sqrt{1-x}\,  \left(1 -\frac{(1+a^2) q^2}{x}\right)^{3/2}},\nn
&=&\frac{(x^2 - q^2)\left[x - (1 + a^2) q^2\right] + a^2 q^2 x (1 - x)}{2\, x^3\, \sqrt{1-x}\,  \left(1 -\frac{(1+a^2) q^2}{x}\right)^{3/2}}, 
\eea
where the last equality indicates $c(x) > 0$ for $x$ in the range of $\sqrt{1+a^2}q<x<1$. The free energy or $\bar I_E$ takes its extremal value, from (\ref{partial action}), when 
$b_{a, q} (\bar x) = \bar b$. Then at $x = \bar x$, we have 
\be
\left.\frac{\partial ^2\bar I_E}{\partial x^2}\right |_{x = \bar x} = - c( \bar x) \left. \frac{\partial b_{a, q} (x)}{\partial x}\right |_{x = \bar x}.
\ee
Since $c (\bar x) > 0$, we have a local minimum of $\bar I_E$ or free energy at $x = \bar x$ if 
\be\label{tcriteria}
 \left.\frac{\partial b_{a, q} (x)}{\partial x}\right |_{x = \bar x}  < 0.
 \ee
 In other words, only when the slope of the inverse reduced temperature function $b_{a, q} (x)$  is negative at the location $ x = \bar x$ with $\bar x$ determined by  $\bar b = b_{a, q} (\bar x)$, the corresponding $\bar I_E$ or the free energy takes its local minimal value, giving a locally (thermally) stable phase.   This is the criterion which we will use to analyze the thermal stability of the black holes considered.

\section{The analysis of phase structures of the stringy black holes}
We are now ready to analyze the thermal phase structures of the 4-dimensional stringy black holes given in \cite{Garfinkle:1990qj}. The $a = 0, \sqrt{3}$ cases have already been analyzed in \cite{Lundgren:2006kt, Lu:2010xt}. In this section, we will discuss the phase structures of black holes for a general dilaton coupling $a$ including the $a = 0, \sqrt{3}$ cases. We will see that the dilaton coupling plays the key role in determining the phase structures of charged black holes. 

\subsection{Chargeless case}
Let us first focus on the chargeless case for a general dilaton coupling $a$.  When the charge $Q$ in (\ref{charge}) is set to vanish, we have $r_- = R_- = 0$ and the only horizon is at
$r_+ = R_+$.  Now the dilaton $\phi$ is just a constant given by its asymptotical value $\phi_0$ and the dilaton coupling $a$ disappears in all the relevant quantities for the present discussion. So the chargeless black hole for a general dilaton coupling $a$ is just the usual Schwarzschild black hole, which can be seen from (\ref{bhsolution}).  We therefore expect the same thermodynamics here as given in \cite{York:1986it, Lundgren:2006kt, Lu:2010xt}.  In other words, there exists a minimal temperature $T_{\rm min}=\frac{3\sqrt{3}}{8\pi r_B}$ below which the only thermally stable phase is the so-called `hot empty space' with the same temperature as preset by the cavity. Above this temperature, we have two solutions from  the thermal equilibrium condition $b_{0} (\bar x) = \bar b$ discussed in the previous section, one corresponding to a small black hole and the other to a large black hole . Only for the large black hole, the free energy is a local minimum.  There exists another so-called transition temperature $T_t =   \frac{27}{32\pi r_B} > T_{\rm min}$ at which the black hole and `hot empty space' can coexist since both of them have same (zero) free energy. For temperature $T_{\rm min} < T < T_t$, the large black hole has a higher free energy than the  corresponding `hot empty space' has, therefore the black hole is only meta-stable and will make a transition to the `hot empty space' via the so-called Hawking-Page transition. Only for temperature $T > T_t$, the large black hole has a negative free energy and becomes the globally stable phase. The detail of this  is referred to \cite{York:1986it, Lundgren:2006kt, Lu:2010xt} and will not be repeated here.

\subsection{Charged case}
In this subsection, we will discuss how the underlying phase structure depends on the reduced charge $q$ for a given dilaton coupling $a$. Note that $a < 0$ case can be accounted for by sending $\phi \to - \phi$. So, without loss of generality, we limit ourselves to $a \ge 0$ in the following discussion. We further sub-divide 
our following discussion to the following three sub-cases: 1) $0 \le a < 1$, 2) $a = 1$, and 3) $a > 1$. 

    As discussed in the previous section, the local stability of a phase is determined by the behavior of $b_{a, q} (x)$ at $x = \bar x$ with $\bar x$ determined by the thermal equilibrium equation $b_{a, q} (\bar x) = \bar b$.  In general, the underlying phase structure is determined by the behavior of $b_{a, q} (x)$ at the lower end of $x$, i.e., $x \to \sqrt{1 + a^2} q$, following the previous study of charged black branes \cite{Lu:2010xt, Lu:2011da}.  We will see that this continues to hold true and in the present case this behavior is completely determined by whether the dilaton coupling $a$ is less or no less than unity.  For $a > 1$, the behavior of $b_{a, q} (x)$ is similar to that of chargeless case and  indeed the underlying phase structure is also characteristically the same as that of chargeless case but with the replacement of `hot empty space' by the corresponding extremal black hole. The $a = 1$ case just resembles the 5-brane case in 10 dimensional string theories given in \cite{Lu:2010xt}. And the $a < 1$ case resembles the p-brane case for $p \le 4$ in 10 dimensional string theories or the M2/M5 brane case in 11 dimensional M-theory \cite{Lu:2010xt}. We will give a detail analysis in what follows. 

For the convenience of discussion, we re-write the function $b_{a, q} (x)$ given in (\ref{bf}) below 
\be \label{bfunction}
b_{a, q} (x) = x \sqrt{1-x} \left(1-\frac{(1+a^2)q^2}{x}\right)^{\frac{1-3a^2}{2(1+a^2)}}\left(1-\frac{(1+a^2)q^2}{x^2}\right)^{\frac{a^2-1}{a^2+1}},
\ee
from which we see that the behavior of $b_{a, q} (x)$ remains the same for $x \to 1$ for all dilaton coupling $a$ but its behavior at the lower end depends crucially on whether the dilation coupling is less than, equal to or greater than unity. 

\noindent
{\bf Case 1: $0 \le a < 1$}

\noindent
For this case, note that $q\sqrt{1 + a^2} < x < 1$. So it is clear from (\ref{bfunction}) that $b_{a, q} (x \to 1) \to 0$ while $b_{a, q} (x \to q\sqrt{1 + a^2}) \to \infty$. This blowing-up behavior of $b_{a, q} (x)$ when $x \to q\sqrt{1 + a^2}$ is due to the last factor on the right of (\ref{bfunction}) for $0\le a < 1$.  This is a typical behavior of the van der Waals 
liquid-gas type phase structure as discussed for $a = 0$ charged black hole in \cite{Lundgren:2006kt} and charged black p-branes in \cite{Lu:2010xt}.  In general, there exists a critical charge $q_c$ which is uniquely determined by $\partial b_{a, q_c} (x)/\partial  x = 0$ and $\partial^2 b_{a, q_c} (x)/\partial  x^2 = 0$ at $x = x_c$.  In other words, at $q = q_c$, $x = x_c$ becomes a reflection point. It is obvious that the critical charge $q_c$ depends on the given dilaton coupling $a$.  Before we discuss how to determine $q_c$, let us give a discussion on the phase structure following 
\cite{Lu:2010xt}.  The phase detail is different for $1/\sqrt{1 + a^2} > q > q_c$ and $0 < q < q_c$.  For this, let us choose a specific $a = 0.2$ for illustration (see Figure 1). 
\begin{figure}
\centering
\subfigure[$q=0.1<q_c$]{\includegraphics[scale=0.5]{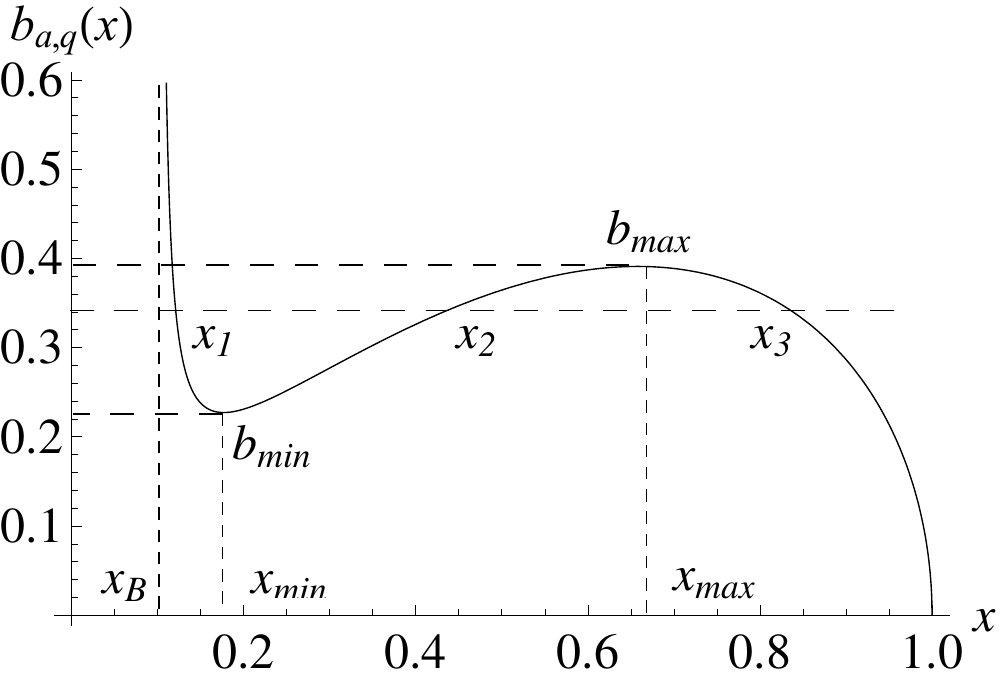}}
\subfigure[$q=0.3>q_c$]{\includegraphics[scale=0.5]{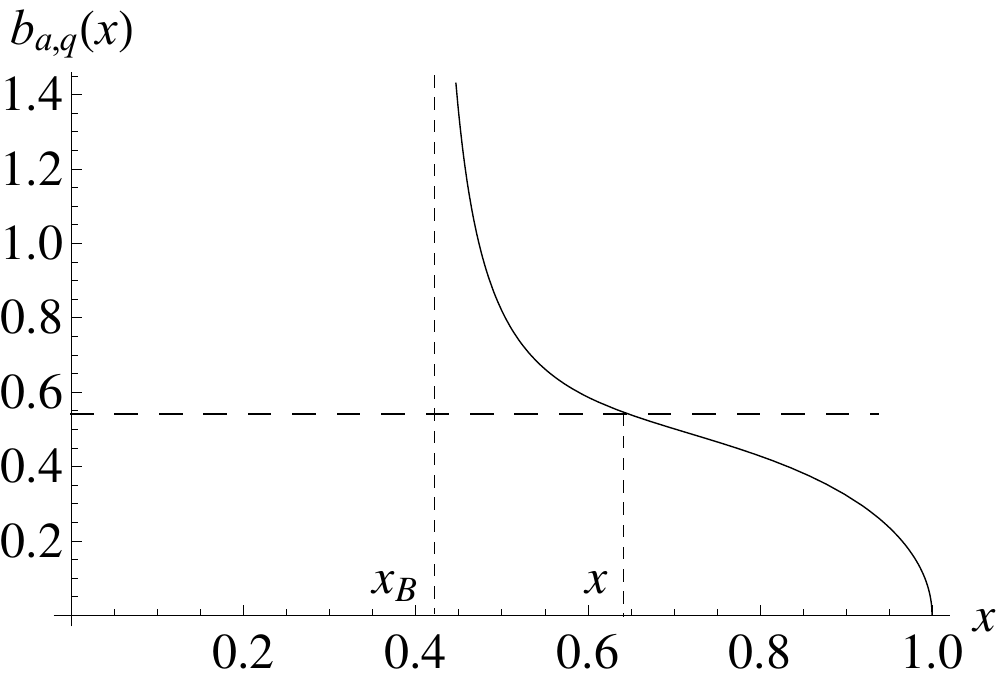}}
\caption{The behaviour of $b(x)$ vs $x$ when $a=0.2$.}
\end{figure}
For this case $q_c \approx 0.24$, when $ q > q_c$, we have a unique solution from $b_{0.2, q} (\bar x) = \bar b$ for each given $\bar b$. It is clear from Figure 1 that $\partial b_{0.2, q} (\bar x)/\partial \bar x < 0$, therefore corresponding to a minimal  free energy. This turns out to be the globally thermally stable black hole configuration. In other words, for $q > q_c$ and for each given $\bar b$, there exists only one thermally stable black hole determined by $b_{a, q} (\bar x) = \bar b$.  For $0 < q < q_c$,  from Figure 1, we can see that $b_{0.2, q} (x)$ has a local minimum $b_{\rm min}$ and a local maximum $b_{\rm max}$.  While there exists a unique solution from $b_{0.2, q} (\bar x) = \bar b$ when $\bar b > b_{\rm max}$
 or $\bar b  < b_{\rm min}$,  there exist in general three solutions denoted as $x_1, x_2, x_3$ (with $x_1 < x_2 < x_3$) when $b_{\rm min} < \bar b < b_{\rm max}$.  The slope of $b_{0.2, q} (x)$ is negative at either $x_1$ or $x_3$ but positive at $x_2$, so the free energy at either $x_1$ or $x_2$ is a local minimum, while it is a local maximum at $x_2$. So only for the small black hole or the large black hole, we can have potentially thermally stable configuration. Following \cite{Lu:2010xt}, we know for each given $q < q_c$ that there exists a transition temperature $T_t$ at which the free energy for the small or large black hole is the same and therefore they can coexist and make free transition between the two. 
Again following \cite{Lu:2010xt}, the transition temperature $T_t$ is determined by requiring that the area above and below the line $b_{0.2, q} (\bar x) = \bar b_t$ and also enclosed by the line  and the curve $b_{0.2, q} (x)$ be the same.  Note that the transition is a first-order one since it involves an entropy change when the large and small black holes make a transition (The entropy depends on the size $x$) .  When $q = q_c$, we have $b_{\rm min} = b_{\rm max} = b_c$ and $x_{\rm min} = x_{\rm max} = x_c$, and there is no entropy change when a transition occurs. So the transition becomes a second-order one and is a critical point.  In other words, when $q < q_c$ moves to $q = q_c$, we have a first-order transition line ending on a second-order critical point.  So overall, this gives a typical van der Waals-Maxwell gas-liquid type phase structure. 

We now come to determine the critical quantities $q_c, x_c$ and $b_c$, which depend on the dilaton coupling $a$ as mentioned earlier. This can be determined by the requirement that $\partial b_{q,a}(x)/\partial x=0$ and $\partial ^2 b_{q,a}(x)/\partial x^2=0$ at $x = x_c$. There exists, however,  a simpler way to determine $q_c$. Following  \cite{Lu:2010xt}, when $q = q_c$, i.e., at the critical point, $x_{\rm min}$ and $x_{\rm max}$ coincide.  We will make use of this to determine $q_c$.  Since $x_{\rm min}$ and $x_{\rm max}$ are determined by
\be \label{extremal}
\frac{\pa b_{a,q}}{\pa x}=\frac{K(x)}{2 x^3\sqrt{1-x} \left(1 - \frac{(1+a^2) q^2}{x}\right)^{\frac{1}{2}+\frac{2 a^2}{1+a^2}} \left(1 - \frac{(1+a^2) q^2}{x^2}\right)^{\frac{2}{1+a^2}}} =0,
\ee
where 
\begin{align}\label{xc}
K(x) = & - 3 \left[x^4 - \frac{2}{3}\left[1 + (1 + 3 a^2) q^2\right] x^3 + 2 q^2 (a^2 - 1) x^2\right. \nn
         & \left.\qquad + \frac{2 q^2}{3}\left[ (3 - a^2)  + (1 + a^2) (3 + a^2) q^2\right] x - \frac{1}{3} (1 + a^2) (5 + a^2) q^4\right].
\end{align}
The denominator on the right side of (\ref{extremal})  is always positive for $x$ in the range of $\sqrt{1+a^2}q<x<1$.  So the extremes amount to setting  $K(x) = 0$. Since $K(x)$ is a quartic polynomial, so the corresponding equation has in general four roots including $x_{\rm min}$ and $x_{\rm max}$.  At the critical point, $x_{min}$ and $x_{max}$ coincide and  the discriminant of the polynomial equation $ K (x) =0$ vanishes. Let us give a detail analysis for the present case.  The discriminant of $K (x) = 0$ is 
\be\label{disc}
  \Delta (a, q) = - 256 q^6 \left[1 - (1 + a^2) q^2\right]^3 h (a, q),
 \ee
  where 
  \begin{align}\label{criticalc}
  h(a, q) =& \left(3 a^4 + 10 a^2 + 3\right)^3 q^6 + 3 \left(9a^{10} + 33 a^8 - 206 a^6 - 150 a^4 -27 a^2 - 171\right) q^4  \nonumber\\ 
   & + 3\left(3 a^8 - 8 a^6 + 114 a^4 - 216 a^2 + 171\right) q^2 + (a^2 - 3)^3 \nn
  = & - \left[(3 - a^2) - \left(3a^4 + 10 a^2 + 3\right) q^2 \right]^3 + 3 \times 12^2 (a^2 - 1)^2 q^2 \left[1 - (1 + a^2) q^2\right],
\end{align}
where the last line is useful for determining the existence of critical charge $q_c$.  We stress that (\ref{xc}), (\ref{disc}) and (\ref{criticalc}) are valid for all $a \ge 0$ and will be used later on for the discussions of $a = 1$ and $a > 1$ cases, respectively.  Given that $0 < q < 1/\sqrt{1 + a^2}$, so $ \Delta (a, q)  = 0$ amounts to setting the above $h (a, q)  =0$.  Let us see if there exists a unique solution in the allowed range for $q$ and for the present case of given $a$ with  $0 \le a < 1$. For this, let us examine the behavior of the first term in the last line of (\ref{criticalc}) first. This term $f (a^2, q^2) = [(3 - a^2) - (3 a^4 + 10 a^2 + 3) q^2]^3$ decreases monotonically with $q^2$ and $f(a^2, 0) = (3 - a^2)^3 > 0$ for $0\le a^2 < 1$. Note that 
\be \label{fterm}
f(a^2, q^2 = 1/(1 + a^2)) = - 2^6 a^6\left(\frac{a^2 + 2}{a^2 + 1}\right)^3 < 0.
\ee  
The second term in the last line of (\ref{criticalc}) is $g (a^2, q^2) = 3 \times 12^2 (a^2 - 1)^2 q^2 \left[1 - (1 + a^2) q^2\right] > 0$ in the range of $0 < q^2 < 1/(1 + a^2)$. We have $g (a^2, 0) = 0$ and $g (a^2, q^2 = 1/(1 + q^2)) = 0$.  Also $g(a^2, q^2)$ has a unique maximum at $q^2 = 1/(2(1 + a^2))$ in the allowed range.  Given (\ref{fterm}),  the curve of $f(a^2, q^2)$ must intersect that of $g(a^2, q^2)$ once and only once in the range of allowed $q^2$. So there exists a unique solution of $h (a^2, q^2) = 0$ or $ \Delta (a, q)  = 0$. Following the discussion given in \cite{Lu:2010xt}, we conclude that the intersection point gives $q = q_c$. Once $q_c$ is determined, we can use (\ref{xc}) to determine $x_c$ and then determine $b_c$ using $ b_c = b_{a, q_c} (x_c)$ for each given $a$ for $0\le a < 1$. 
Some sample values of $q_c, x_c, b_c$ for different values of $a$ are given in Table 1.
  \begin{table}[!h]
        \centering 
        \begin{tabular}{|c|c|c|c|}
        \hline
        a&$q_c$&$x_c$&$b_c$\\ \cline{1-1}\cline{1-2}\cline{1-3}\cline{1-4}
        0&0.236068 &0.527864&0.428784\quad\\ \cline{1-1}\cline{1-2}\cline{1-3}\cline{1-4}
        0.2&0.237828&0.528475&0.429447\\ \cline{1-2}\cline{1-3}\cline{1-4}
        0.4&0.243500&0.530378&0.431635\\ \cline{1-2}\cline{1-3}\cline{1-4}
        0.6&0.254610&0.533810&0.436145\\ \cline{1-2}\cline{1-3}\cline{1-4}
        0.8&0.275912&0.539120&0.445739\\ 
  \cline{1-2}\cline{1-3}\cline{1-4}
        $\rightarrow$ 1&0.353553 &0.5&0.5\\ \cline{1-2}\cline{1-3}\cline{1-4}
        \hline
        \end{tabular}
        \centerline{}
        \centerline{Table 1: Some sample values of $q_c, x_c, b_c$ for different values of $0  < a < 1$.} 
         \end{table}
The critical charge $q_c$ and $x_c, b_c$ given in the table for $a = 0$ all agree with the previous given values in  \cite{Lundgren:2006kt,Lu:2013nt}.   We shall argue that $q_c$  approaches $\sqrt{2}/4$ when  $a \to 1$ in the following $a = 1$ case.   

\noindent
{\bf Case 2: $a=1$}

By setting $a=1$ in \eqref{bfunction},  we have
\be
b_{1,q}(x)=\frac{x \sqrt{1-x}}{\sqrt{1-2q^2/x}}.
\ee
Recall that $\sqrt{2} q < x < 1$, we have from the above that $b_{1, q} (x \to 1) \to 0$ while $b_{1, q} (x \to \sqrt{2} q) \to \sqrt{2} q$.  To understand the behavior of $b_{1, q} (x)$ in the allowed range of $x$, we first examine its possible extremes. For this,  we have
\be
\frac{\partial b_{q,1}(x)}{\partial x}=c(x)\left(-3x^2+2(1+4q^2)x-6q^2\right)=c(x)B(x,q)
\ee
where
\be \label{B}
c(x)= \frac{1}{2 x \left(1-\frac{2 q^2}{x}\right)^{3/2}  \sqrt{1-x}}, \qquad B (x) = - 3\left[x^2 -  \frac{2}{3} (1+4q^2)x + 2q^2\right] .
\ee
In the above,  $c(x)$ is always positive in the allowed range of $x$. So $\partial b_{1,q} (x)/\partial x = 0$ amounts to setting $B (x) = 0$, i.e.,
\be \label{bsolution}
x^2 -  \frac{2}{3} (1+4q^2)x + 2q^2 = 0,
 \ee
from which we have two solutions 
\be\label{tworoots}
x_\pm = \frac{1}{3} \left[ (1 + 4 q^2) \pm \sqrt{(1 - 2 q^2)(1 - 8 q^2)}\right],
\ee
provided $q \le 1/2\sqrt{2}$ (due to  $\sqrt{2} q < 1$).  One can check easily that $x_- < \sqrt{2} q$ and $\sqrt{2} q < x_+ < 1$ for $q < 1/2\sqrt{2}$. So $x_-$ is not in the allowed region of $\sqrt{2} q < x < 1$, therefore excluded.  Only $x_+$ is  and so for $q < 1/2\sqrt{2}$, we have one extreme in the allowed range of $x$.  We will comment $x_-$ later on when $q = 1/2\sqrt{2}$. Now let us examine further the exact nature of this extreme at $x_+$.  For this, we have
\be
\left.\frac{\partial B (x)}{\partial x}\right |_{x = x_+}  = - 6 \left(x_+ -  \frac{1 + 4 q^2}{3}\right)  < 0,
\ee
which implies that $x = x_+$ gives a maximum of $b_{1, q}$ when $q < 1/2\sqrt{2}$.  This further implies that $b_{1, q} (x)$ increase with $x$ in the range of $\sqrt{2} q < x < x_+$, takes its maximal value at $x = x_+$ as
\be\label{bmax}
b_{\rm max} = \frac{1}{9} \left[4 q^2 + 1 + \sqrt{(1 - 2 q^2)(1 - 8 q^2)}\right] \left[2 \sqrt{1 - 2 q^2} - \sqrt{1 - 8 q^2}\right]^2,
\ee
then decreases for x in the range of $x_+ < x < 1$. The characteristic behavior of this is given in the first graph of Figure 2 for the case of $q = 0. 2 < 1/2\sqrt{2}$. 
\begin{figure}[!h]
\centering
\subfigure[$q=0.2$]{\includegraphics[scale=0.5]{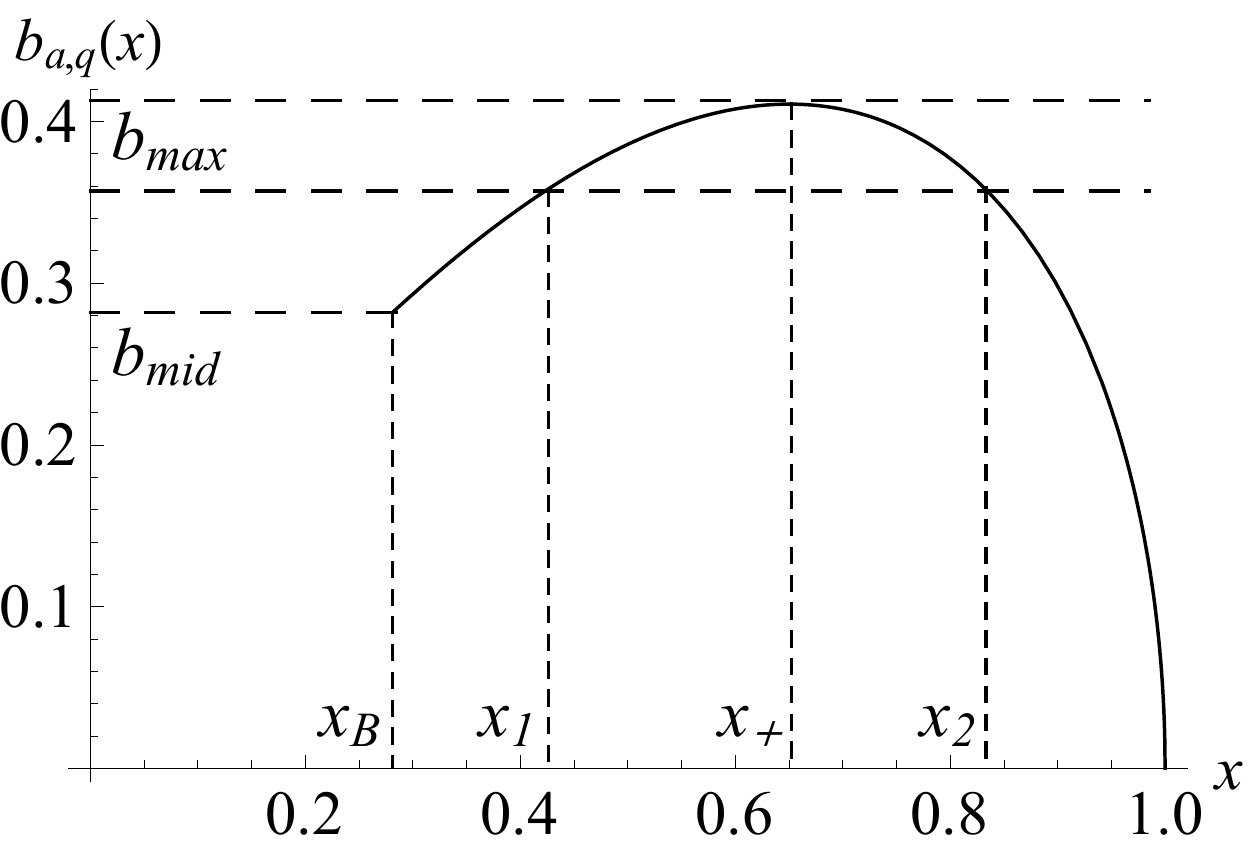}}
\subfigure[$q=0.5$]{\includegraphics[scale=0.5]{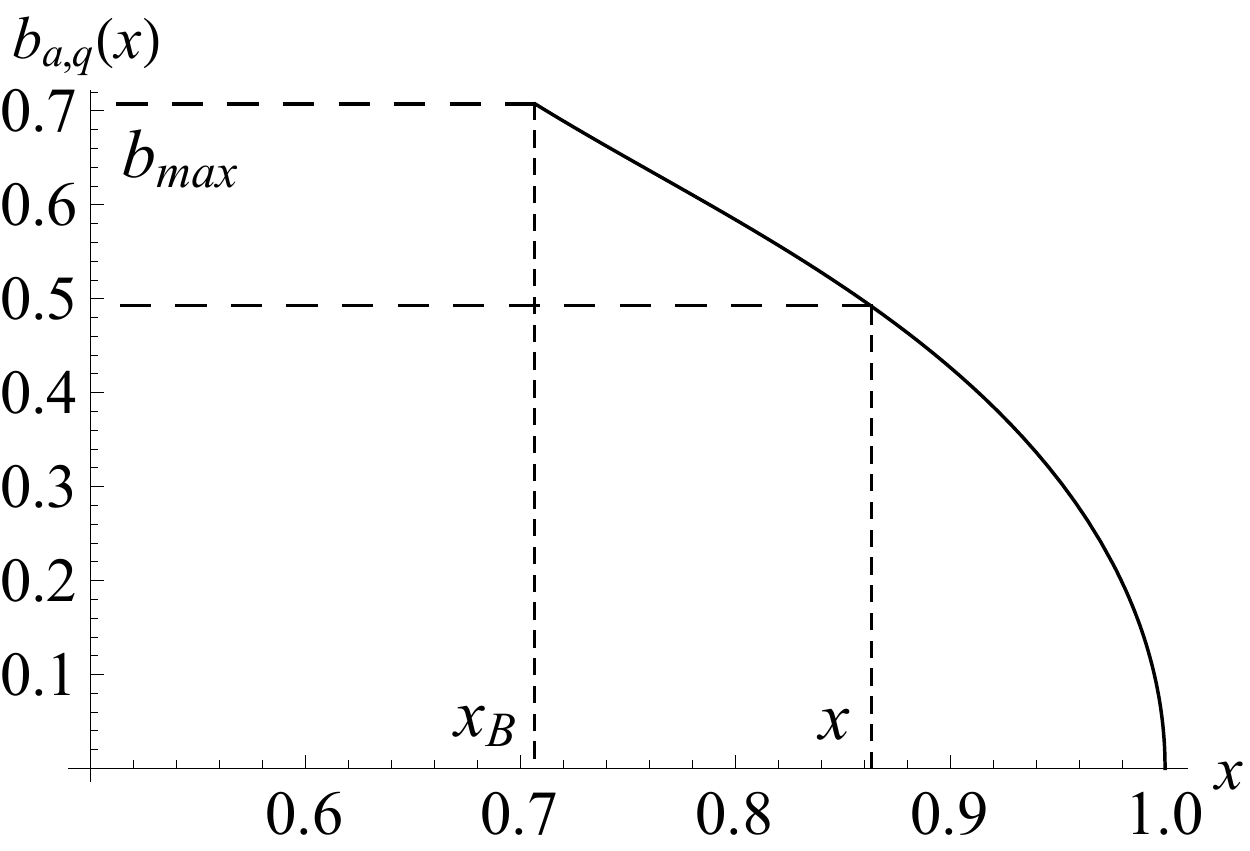}}
\caption{The behaviour of $b(x)$ vs $x$ when $a=1$. Here $x_B=\sqrt{2}q$}
\end{figure}
Let us now focus on the case $q = q_c = 1/2\sqrt{2} = \sqrt{2}/4$. One can now read from (\ref{tworoots}) that $x_- = x_+ = x_c = 1/2 = \sqrt{2} q_c$ ($b_c = 1/2$). This is a reflection point but unlike the case of $0 < a < 1$ discussed previously, this point is not a critical one since it is at the lower end of $x$. For this reason,  there is no van der Waals liquid-gas type phase structure discussed previously.  This also answers $q_c \to \sqrt{2}/4$ as $a \to 1$ raised in the case of $0 < a < 1$.  Note that this $q_c = 1/2\sqrt{2}$ can also be easily determined from the $h (a^2, q^2) $ given in (\ref{criticalc}) when $a = 1$ for which the second term in the last line vanishes.   Now for $1/2\sqrt{2} < q < 1/\sqrt{2}$,  the equation (\ref{bsolution}) has no real roots. Then the sign of $B (x)$ in the range of $\sqrt{2} q < x < 1$ will determine whether $b_{1, q} (x)$ is a monotonically increasing or decreasing function.  For this,  we have, from (\ref{B}),
\bea 
B (\sqrt{2} q) &=& 8\sqrt{2} q \left(q - \frac{1}{\sqrt{2}}\right)\left(q - \frac{1}{2\sqrt{2}}\right) < 0,\nn
B(1) &=& - (1 - 2 q^2) < 0,
\eea 
which must therefore imply that $b_{1, q} (x)$ is a monotonically decreasing function in the range of $\sqrt{2} q < x < 1$ for $1/2\sqrt{2} < q < 1/\sqrt{2}$.  The typical behavior of this is given in the second graph in Figure 2 for $q = 0.5$. 

Given what has been said above, the phase structure for this case is characteristically the same as that for the $\tilde d = 2$ case for black p-brane discussed in \cite{Lu:2010xt, Lu:2012rm}. There are three sub-cases, depending on $0 < q < q_c = 1/2\sqrt{2}, q = q_c$ and $q_c < q < 1/\sqrt{2}$, respectively, but there are no van der Waals-Maxwell gas-liquid type phase structure and the second-order phase transition at $q = q_c$. Each of the sub-cases is just like the chargeless case but with the replacement of `hot empty place' with the corresponding extremal black hole as discussed in \cite{Lu:2011da} for the $\tilde d = 2$ case.  For example, when $0 < q < q_c $, for $\bar b > b_{\rm max}$, the only stable phase is the corresponding extremal black hole with the preset $\bar b$. When $b_{\rm mid} \le \bar b < b_{\rm max}$, only the large black hole is locally stable. For the large black hole, we need to compare its free energy with that of the corresponding extremal black hole. If its free energy is larger, then a Hawking-Page transition will take place and the globally stable phase is the extremal black hole. Only for the large black hole with its free energy equal to or smaller than the corresponding extremal black hole can become globally stable one.  The detail can be similarly discussed as in the chargeless case or is referred to \cite{Lu:2010xt, Lu:2011da}.  We will not repeat it here. 

\noindent
{\bf Case 3: $a>1$}

This is the case, just like that of  charged black p-branes with $p = D - 4$ in D dimensions studied in \cite{Lu:2010xt} (one example is $p = 6$ in 10 dimensional Type IIA theory), giving a phase structure which resembles that of chargeless case discussed earlier but with the replacement of `hot empty space' by the extremal black hole as discussed in \cite{Lu:2010xt, Lu:2011da}.  For this case, we have first from (\ref{bfunction})
\be \label{bfunc}
b_{a, q} (x) = \frac{x \sqrt{1-x} \left(1-\frac{(1+a^2)q^2}{x^2}\right)^{\frac{a^2-1}{a^2+1}}}{\left(1-\frac{(1+a^2)q^2}{x}\right)^{\frac{3a^2 - 1}{2(1+a^2)}}},
\ee
which gives  $b_{a, q} (x\to q\sqrt{1 + a^2}) \to 0$ and $b_{a, q} (x \to 1) \to 0$. This is exactly the behavior of charged black p-branes with $p = D - 4$ in D dimensions mentioned above.  At a first look, due to the presence of dilaton coupling $a$, one may suspect that the behavior of  above $b_{a, q} (x)$ could be 
different from that of charged black (D - 4)-branes for some $a$  within the range $\sqrt{1 + a^2} q < x < 1$. For example, there might be a possibility of appearance of two maxima and a minimum of $b_{a, q} (x)$ in the allowed range of $x$ for suitable parameter $a$. If this could happen, we would have a new phase structure. Unfortunately, as we will demonstrate in what follows, this doesn't  and there is only one extreme of $b_{a, q} (x)$ which is a maximum in the allowed region of $x$, exactly like in the case of charged black (D - 4)-branes in D-dimensions. 

Note that $b_{a, q} (x) > 0$ for $\sqrt{1 + a^2} q < x < 1$ and $\partial b_{a, q} (x) /\partial x = 0$, from (\ref{extremal}) and (\ref{xc}), amounts to the following quartic equation, \bea \label{discnew}
&&x^4 - \frac{2}{3}\left[1 + (1 + 3 a^2) q^2\right] x^3 + 2 q^2 (a^2 - 1) x^2\nn
&& \qquad + \frac{2 q^2}{3}\left[ (3 - a^2)  + (1 + a^2) (3 + a^2) q^2\right] x - \frac{1}{3} (1 + a^2) (5 + a^2) q^4 = 0,
\eea
which has four roots in general. Since the last term in the above is negative, there is at least one negative real root. So we have at most three positive real solutions from the above equation.  We then expect two possibilities\footnote{Since $K (x = q\sqrt{1 + a^2}) = 4 (a^2 - 1) q^3 \sqrt{1 + a^2} \left[1 - q \sqrt{1 + a^2} \left(1 - q\sqrt{1 + a^2}\right)\right] > 0$ for $a > 1$ and $K (x = 1) = - \left[1 - q^2 (1 + a^2)\right]^2 < 0$ with $K (x)$ given in (\ref{xc}), we cannot have the other possibility that $b_{a, q} (x)$ has either a minimum or two minima and one maximum since this function is an increasing function at the lower end while a decreasing one at the other end.}  that $b_{a, q} (x)$ has either one maximum or two maxima and one minimum in the region of $\sqrt{1 + a^2} q < x < 1$.  In the former case, the phase structure is just like that of the charged black (D - 4)-brane case mentioned above.   If the latter were true,  there would have been an interesting phase structure with a second-order critical point.  To see whether the latter case could happen, let us examine the discriminant of the above equation (\ref{discnew}), which, up to a factor of $3^{- 6}$, is given by (\ref{disc}).  The key for this is the function $h(a, q)$ given in (\ref{criticalc}) and we re-write it here for convenience as
\be 
  h(a, q) = - f (a^2, q^2) +  g (a^2, q^2) ,
\ee
where as given before
\bea\label{fg}
 f(a^2, q^2) &\equiv& [(3 - a^2) - \left(3a^4 + 10 a^2 + 3\right) q^2]^3, \\
 g(a^2, q^2) &\equiv & 3 \times 12^2 (a^2 - 1)^2 q^2 \left[1 - (1 + a^2) q^2\right].
 \eea
 In the above,  $f (a^2, q^2)$ is a monotonically decreasing function of $q^2$ and $f(a^2, 0) = (3 - a^2)^3 >  0$ for $1 < a^2 < 3$, $f(a^2, 0) = 0$ for $a^2 = 3$ and $f(a^2, 0) < 0 $ for $a^2 > 3$, respectively.  Also $g(a^2, 0) = 0$, $g(a^2, q^2 = 1/(1 + a^2)) = 0$. Further  $g(a^2, q^2) > 0$ for $0 < q^2 < 1/(1 + a^2)$ and has a maximum at $q^2 = 1/(2(1 + a^2))$. Since $f (a^2, q^2) < 0$ at the upper end of $q^2$ from (\ref{fterm}) when $a^2 \ge 3$,  therefore the curve of $f(a^2, q^2)$ can never intersect that of $g(a^2, q^2)$ for any $q > 0$  in the region of $ 0 < q < 1/\sqrt{1 + a^2}$ for this case.  For $1 < a^2 < 3$, the curve of $f (a^2, q^2)$ does intersect  that of $g(a^2, q^2)$ in the allowed range of $q^2$ once and only once. So one expects a unique solution of $h (a^2, q^2) = 0$ for $q_c$.  This appears as a critical charge.  However, as we will show in what follows,  the corresponding $x_c < q_c \sqrt{1 + a^2}$, therefore not in the allowed range of $x$. So we have only the largest root to be the correct one. In other words, in the allowed range of $x$, we have again only one maximum.  For the present situation, the characteristic picture of the 4 roots of (\ref{discnew}) is shown in Figure 3. Let us set $F (x)$ equal the left side of (\ref{discnew})
 \bea
&& F (x) \equiv x^4 - \frac{2}{3}\left[1 + (1 + 3 a^2) q^2\right] x^3 + 2 q^2 (a^2 - 1) x^2 \nn
&&\qquad \qquad + \frac{2 q^2}{3}\left[ (3 - a^2)  + (1 + a^2) (3 + a^2) q^2\right] x - \frac{1}{3} (1 + a^2) (5 + a^2) q^4,
\eea
 from which 
 \be 
 F' (x) = 4 x^3 - 2 \left[1 + (1 + 3 a^2) q^2\right] x^2 + 4 q^2 (a^2 - 1) x  + \frac{2 q^2}{3}\left[ (3 - a^2)  + (1 + a^2) (3 + a^2) q^2\right],
 \ee
 where $F' (x) = d F(x)/d x$. We then have for $1 < a^2 < 3$,  
 \be 
 F (0) = - \frac{1}{3} (1 + a^2) (5 + a^2) q^4 < 0, \quad F' (0) = \frac{2 q^2}{3} \left[ (3 - a^2) + (1 + a^2) (3 + a^2) q^2\right] > 0,
 \ee
 and at the lower end of the allowed range of $x$, i.e., $x  = q \sqrt{1 + a^2}$, 
 \bea 
 F (q \sqrt{1 + a^2}) &=&  \frac{4 q^3}{3} (1 - a^2) \sqrt{1 + a^2} \left(1 - q\sqrt{1 + a^2}\right)^2 < 0, \nn
  F' (q \sqrt{1 + a^2}) &=& - \frac{8 a^2 q^2}{3} \left(1 - q\sqrt{1 + a^2}\right) \left(1 - 2 q \sqrt{1 + a^2}\right).
 \eea 
 Given that the present $b_{a, q} (x)$ has at least one maximum in the allowed range of $x$ and $F(x) > 0, F' (x) > 0$ for sufficient large $x$, there must exist a maximum of $F (x)$ in the range of $0 < x < q \sqrt{1 + a^2}$ since $F (0) < 0, F' (0) > 0$ and $F(q \sqrt{1 + a^2}) < 0$ with $F' (q \sqrt{1 + a^2}) \le 0$ when $ q \le 1/(2\sqrt{1 + a^2})$ and $F' (q\sqrt{1 + a^2}) > 0$ when $1/\sqrt{1 + a^2} > q > 1/(2\sqrt{1 + a^2})$, noting that $x = q\sqrt{1 + a^2}$, the lower end of allowed $x$, moves to its large value, i.e., towards the right side,  rather than to its small value, when $q$ increases for fixed $a$.  In other words, the lower end of allowed $x$ must be the one as shown in Figure 3.  
  \begin{figure}[!h]
\centering
\includegraphics[scale=0.5]{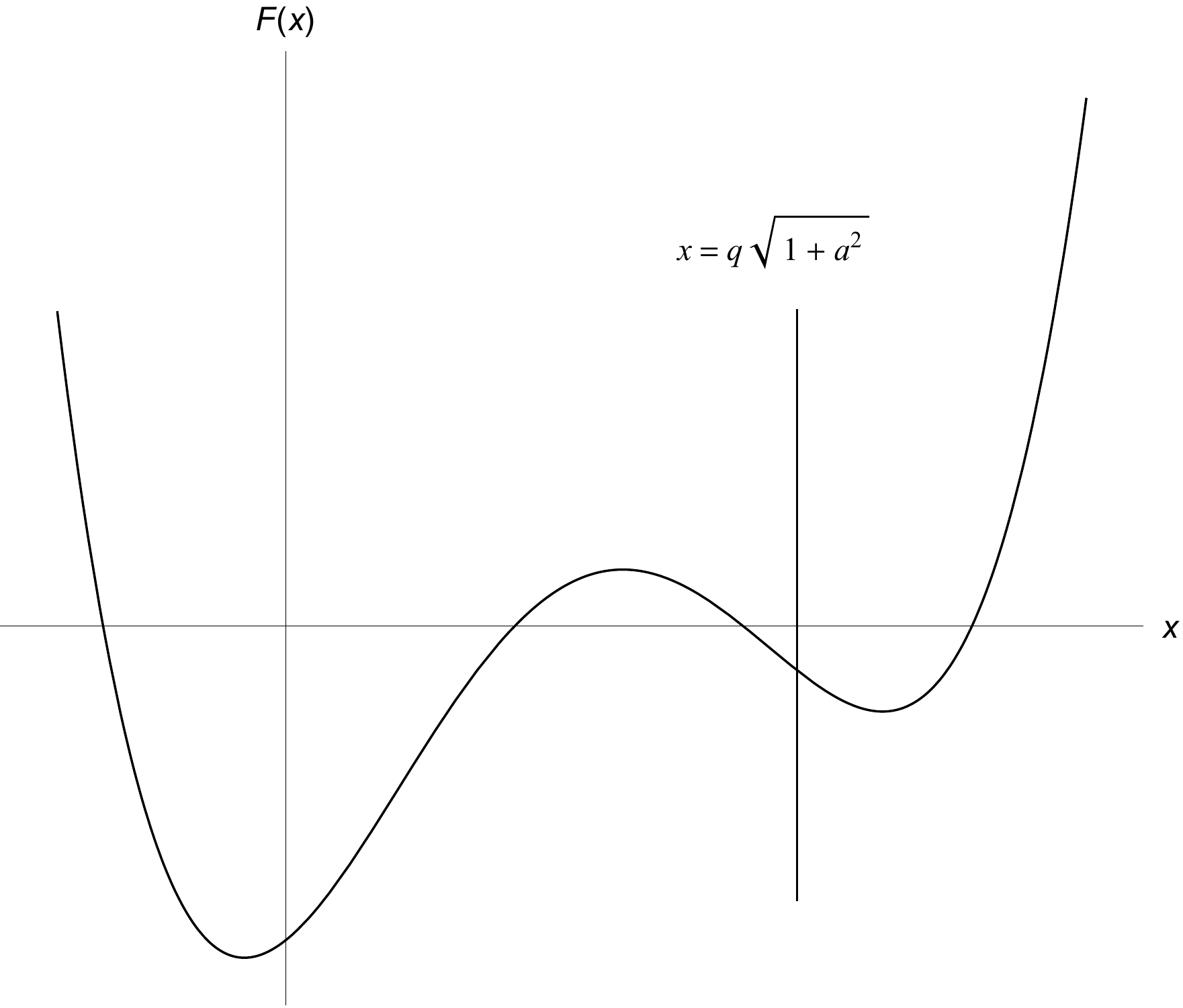}
\caption{The characteristic behavior of $F (x) $ vs $x$ for  $1< a^2 < 3$.}
\end{figure}
 This says that we can have only one maximum for $b_{a, q} (x)$ even in this case and therefore there is no new phase structure. We also have numerical tests for this.  When $q = q_c$, determined from $h (a^2, q^2) = 0$, for a choice of $a^2$ in the region of $1< a^2 < 3$, there are one negative root, two identical positive roots $x_c$ and one positive largest root $x_{\rm max}$ from (\ref{discnew}).   This $q_c$ indeed gives an $x_c > 0$ but this $x_c$ is smaller than the value of $q_c \sqrt{1 + a^2}$, therefore outside the allowed region of $x$.  So only $x_{\rm max}$ is the wanted one, giving rise to a maximum of $b_{a, q} (x)$.  So we don't have a critical point for this case, either.  A few samples of numerical computations for $q_c, x_c$ and $q_c \sqrt{1 + a^2}$ are given in Table 2.   
   \begin{table}[!h]
        \centering 
        \begin{tabular}{|c|c|c|c|}
        \hline
        $a^2$&$q_c$&$x_c$&$q_c\sqrt{1 + a^2}$\\ \cline{1-1}\cline{1-2}\cline{1-3}\cline{1-4}
        1.2&0.2161&0.2260&0.3205\\ \cline{1-1}\cline{1-2}\cline{1-3}\cline{1-4}
        1.4&0.1455&0.1341&0.2254\\ \cline{1-2}\cline{1-3}\cline{1-4}
        1.6&0.0988&0.0811&0.1593\\ \cline{1-2}\cline{1-3}\cline{1-4}
        1.8&0.0665&0.0486&0.1112\\ \cline{1-2}\cline{1-3}\cline{1-4}
        2.0&0.0435&0.0281&0.0754\\ \cline{1-2}\cline{1-3}\cline{1-4}   
        2.2&0.0272&0.0153&0.0486\\ \cline{1-2}\cline{1-3}\cline{1-4}  
        2.4&0.0156&0.0074&0.0287\\ \cline{1-2}\cline{1-3}\cline{1-4}
        2.6&0.0075&0.0029&0.0143\\ \cline{1-2}\cline{1-3}\cline{1-4}          
        2.8&0.0024&0.0006&0.0046\\\cline{1-2}\cline{1-3}\cline{1-4}
        \hline
        \end{tabular}
        \centerline{}
        \centerline{Table 2: Some sample computations of $q_c, x_c, q_c \sqrt{1 + a^2}$ for different values of $1 < a^2 < 3$.} 
         \end{table}
 \begin{figure}[!h]
\centering
\subfigure[$a^2 = 2$]{\includegraphics[scale=0.5]{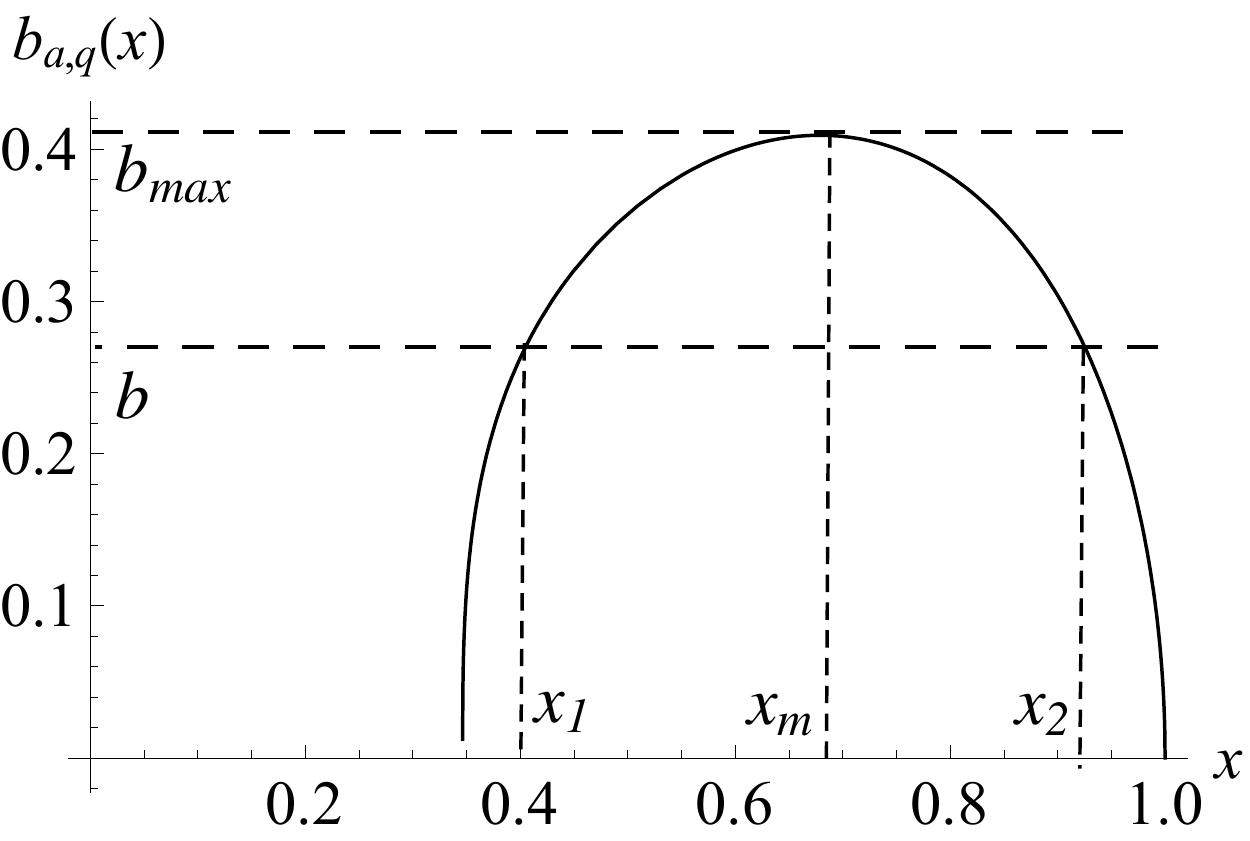}}
\subfigure[$a^2 = 4$]{\includegraphics[scale=0.6]{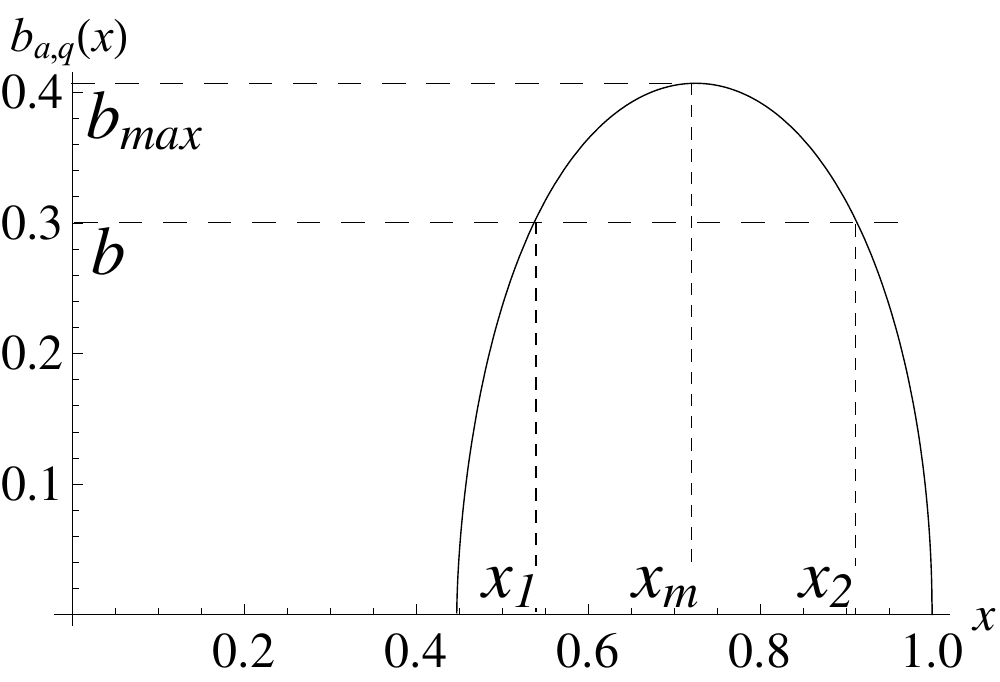}}
\caption{The behaviour of $b_{a, q}(x)$ vs $x$ with $q = 0.2$ for  $a^2=2, 4$, respectively.}
\end{figure}
So when $a^2 > 1$,  $b_{a, q} (x)$ has an unique maximum in the region of $q \sqrt{1 + a^2} < x < 1$ and the corresponding phase structure is characteristically the same as that of charged black (D - 4)-branes in D-dimensions . Figure 4 gives an illustration of $a^2 = 2$ and $a^2 = 4$ cases, respectively.

\section{Connection to the higher dimensional branes}
It is well-known that the 4-dimensional black holes with different dilaton coupling $a$ can be obtained from higher dimensional p-branes and their bound states/intersecting branes via dimensional reductions (direct and/or diagonal (double) dimensional reductions) \cite{Lu:1995cs, Lu:1995yn,Papadopoulos:1996uq, Tseytlin:1996bh, Klebanov:1996mh, Behrndt:1996pm,Gauntlett:1996pb,Khviengia:1996mm,Lu:1996mg,Bergshoeff:1996rn, Lu:1997hb}.  Only for $a = 0, 1/\sqrt{3}, 1, \sqrt{3}$, the corresponding extremal black holes are supersymmetric and the $a = 0, 1/\sqrt{3},  1$ extremal ones are the bound states of 4, 3, and 2  extremal $a =\sqrt{3}$ black holes, respectively\cite{Duff:1994jr,Sen:1994eb,Sen:1995in, Sen:1995vr,Duff:1995sm,Cvetic:1995yq,Rahmfeld:1995fm}. These bound states have their intersecting brane interpretations of 4, 3, 2 simple branes in 10 and 11 dimensions\cite{Papadopoulos:1996uq, Tseytlin:1996bh, Klebanov:1996mh,Behrndt:1996pm,Gauntlett:1996pb,Khviengia:1996mm, Bergshoeff:1996rn, Lu:1997hb}. Other values of $a$ are due to non-supersymmetric branes (intersecting branes) in higher dimensions \cite{Lu:1995yn, Lu:1997hb, Gibbons:1994vm}.   

The natural question in the present context is how to relate the phase structures of the 4-dimensional charged black holes with different dilaton coupling $a$ to those of  
their higher dimensional branes or higher dimensional intersecting branes.  We here try to make this connection based on known knowledge of
the phase structures of  the higher dimensional charged black branes \cite{Lu:2010xt, Lu:2011da,Lu:2012rm, Lu:2013nt, Lu:2014dra} and the above mentioned connection of the 4-dimensional black holes to their higher dimensional brane systems.

For this purpose,  the higher dimensional brane systems considered should have at least three overall transverse directions such that they are asymptotically Minkowski, and the corresponding spherical symmetry in the transverse directions, the basic requirements of the present approach.  This serves also to give 4-dimensional black holes when dimensionally reduced to 4 dimensions.    

Let us first consider the case of simple charged black branes in diverse dimensions.  We have knowledge of the phase structures of these systems \cite{Lu:2010xt} with the following dilaton coupling,
 \be \label{dc}
 a_D^2 (p)  = \Delta - \frac{2 d \tilde d}{D - 2},
 \ee
where $d =  p + 1$ and $d + \tilde d = D - 2$ with $D$ the bulk spacetime dimensions.  In this formula,  $\Delta = 4$ for the simple p-branes. In general, supersymmetric p-branes in diverse dimensions can have $\Delta = 4/N$ with $N = 1, 2, 3, 4$ with  N field strengths participating in the solution.  Non-supersymmetric 
extremal or non-extremal p-branes are also possible with many choices of values of $\Delta$ given in \cite{Lu:1995yn, Lu:1997hb, Gibbons:1994vm}. One of important properties in (direct or diagonal (double)) dimensional reduction is that the value of $\Delta$ remains unchanged \cite{Lu:1995cs}.  Given this, for a diagonal (double) dimensional reduction from $(D + 1, p + 1) \to (D, p)$, we have the relation between the $\hat a^2_{D + 1} (p + 1)$ and $a^2_D (p)$ as \cite{Lu:1995cs}
\be\label{ddr}
 \hat a^2_{D + 1} (p + 1) = a^2_D (p) - \frac{2 \tilde d^2}{(D - 2) (D - 1)},
 \ee
 where $\hat a_{D + 1} (p + 1)$ and $a_{D} (p)$ denote the dilaton coupling for $(p + 1)$-brane in $(D + 1)$ dimensions and $p$-brane in D dimensions, respectively. 
 For a direct dimensional reduction from $(\tilde D + 1,  p) \to (\tilde D,  p)$, we have
 \be\label{sdr}
 \tilde a^2_{\tilde D + 1} (p) = a^2_{\tilde D} (p) - \frac{2 d^2}{(\tilde D - 2)(\tilde D - 1)}.
 \ee
 Note that  $\tilde d$ remains unchanged in a diagonal  (double) dimensional reduction while $d = p + 1$ remains so in a direct dimensional reduction.  From (\ref{ddr}), we can obtain the formula for diagonal (double) dimensional reduction k times as
 \be\label{kddr}
 \hat a^2_{D + 1} (p + 1) = a^2_{D + 1 - k} (p + 1 - k) - \frac{2 k \tilde d^2}{(D - 1 - k) (D - 1)},
 \ee 
 from which we can read $a^2_{D - p} (0)$ for $k = p + 1$ as
 \be\label{pddr}
 a^2_{D - p} (0) =  \hat a^2_{D + 1} (p + 1)  +  \frac{2 (p + 1) \tilde d^2}{(D - p -2) (D - 1)}. 
 \ee
 Similarly, from (\ref{sdr}), we can have the following formula for direct dimensional reduction k times as
 \be \label{ksdr}
 \tilde a^2_{\tilde D + 1} (p) = a^2_{\tilde D + 1 - k} (p) - \frac{2 k d^2}{(\tilde D - 1- k)(\tilde D - 1)},
 \ee
 from which we can read for $k = \tilde D - 3, p = 0$ as
 \be\label{40}
 a^2_4 (0) = \tilde a^2_{\tilde D + 1} (0) + \frac{ (\tilde D - 3) }{(\tilde D - 1)} ,
 \ee
 where we have set $d = p + 1 = 1$. Note that $a^2_{D - p} (0) = \tilde a^2_{\tilde D + 1} (0)$ if  $\tilde D  = D - p - 1$. With this and using (\ref{40}) and (\ref{pddr}), we end up with 
 \be\label{pto0}
 a^2_4 (0) =  \hat a^2_{D + 1} (p + 1)  +  \frac{2 (p + 1) (D - p - 3)^2}{(D - p -2) (D - 1)} + \frac{ D - p - 4}{(D- p - 2)},
 \ee
 where we have used $\tilde d = D - 3 - p$.  This is the formula relating the dilaton coupling $\hat a_{D + 1} (p + 1)$ for  $(p + 1)$-branes in $(D + 1)$ dimensions to that for a 0-brane (or black hole) in 4 dimensions via first $(p + 1)$ diagonal (double) dimensional reductions to give 0-branes in $(D - p)$ dimensions and then from the 0-branes to 4-dimensional 0-branes by $(D - p - 4)$ direct dimensional reductions if $D - p - 4 \ge 0$.  Now using (\ref{dc}) for $\hat a_{D + 1} (p + 1)$ in the above, we end up
 with a very simple formula
 \be\label{4dc}
 a^2_4 (0) = \Delta - 1,
\ee 
independent of which $p$ and $D$ we start with.  If we take $D = 4, d = 1$ (therefore $\tilde d = 1$) in (\ref{dc}), we end up also with the same formula as in (\ref{4dc}).  In other words, the 4-dimensional dilaton coupling $a_4 (0)$ is just the one obtained from a higher dimensional $a_D (p)$ if the p-brane in D dimensions can be related to 4-dimensional 0-brane via diagonal (double) plus possible direct dimensional reductions.  This remains true for any lower dimensional dilaton coupling $a_D (p)$ if it can be related to a higher dimensional $a_{D'} (p')$ with  $D' \ge D,\, p' \ge p$ via diagonal (double) plus possible direct dimensional reductions.  This is independent of actual reduction path even though in the above we choose a particular one for illustration.    

With the above preparation, we are now ready to discuss whether the phase structures of higher dimensional charged black p-branes can be preserved when reduced to 4 dimensions. For this, let us consider specifically, for examples, the simple 10-dimensional charged black p-branes considered in \cite{Lu:2010xt} .  For these p-branes, we know that the charged black p-branes have a phase structure of the van der Waals-Maxwell liquid-gas type for $0\le p \le 4$, a phase for $p = 5$ with the appearance of a critical charge but no van der Waals-Maxwell gas-liquid type phase structure, i.e. the one on borderline between the $p \le 4$  case and the $p = 6$ case,   and a phase structure for $p =6$ resembling that of the chargeless case but with the replacement of `hot empty space' by the corresponding extremal 6-brane. For all these branes,  $\Delta = 4$.  

There are no issues for diagonal (double) dimensional reductions of these charged black p-branes but direct dimensional reductions of non-extremal charged black p-branes are not so simple \cite{Myers:1986rx,Lu:1996kg, Cvetic:1996gq, Lavrinenko:1997rc}.  Nevertheless, there are no issues for direct dimensional reductions of extremal charged black p-branes for which  $\tilde d \to \tilde d - 1$ for each direct dimensional reduction while $d = p + 1$ remains unchanged\footnote{This is actually what we need to discuss the phase structure since this structure itself is determined by the behavior of  $b_{a, q} (x)$ at the lower end $x \to q \sqrt{1 + a^2}$, the extremal limit, for the present charged black holes or  that of the $b(x, q)$ for charged black p-branes considered in \cite{Lu:2010xt} at the lower end $x \to q$, also the extremal limit.}. Note that $\Delta$ remains the same in either of these two reductions. So $\Delta = 4$ implies that  the 4-dimensional charged black holes obtained from 10-dimensional p-branes with $0\le p \le 6$ via diagonal (double) and/or direct dimensional reductions have the same dilaton coupling, from (\ref{4dc}), $a^2 = a^2_4 (0) = 3$, giving the same phase structure as the charged black 6-branes as discussed earlier in this paper.  Among these p-branes, only the 6-branes are reduced to 4-dimensional charged black holes via only diagonal (double) dimensional reductions for which  $\tilde d = 1$ remains unchanged during the reductions. This very fact can also be used to understand why the phase structure can be preserved by diagonal (double) dimensional reductions.  Recall that the phase structure itself is determined by the behavior of the $b (x, q)$ function given in \cite{Lu:2010xt}  at the lower end $x \to q$, the extremal limit.  This $b (x, q)$ \cite{Lu:2010xt} is 
\be
b (x, q) = \frac{1}{\tilde d} \frac{x^{1/\tilde d} (1 - x)^{1/2}}{\left(1 -\frac{q^2}{x^2}\right)^{\frac{\tilde d - 2}{2\tilde d}} \left(1 -\frac{q^2}{x}\right)^{\frac{1}{\tilde d}}}, 
 \ee 
whose lower end, i.e., $x \to q$, behavior is solely determined by $\tilde d$, not the bulk spacetime dimension $D$.  In diagonal (double) dimensional reductions, $\tilde d$ remains unchanged, so is the phase structure. For example, for both the 10-dimensional 6-branes and the 4-dimensional charged black holes, they all have the same $\tilde d = 1$, therefore the same phase structure.  However, for direct dimensional reductions, each of these will reduce $\tilde d$ by one, i.e., $\tilde d \to \tilde d - 1$. For examples, for the 10-dimensional 0-branes to give 4-dimensional black holes, we need to have 6 step by step direct dimensional reductions. In other words, we need to have $\tilde d = 7 \to \tilde d - 6 = 1$, therefore ending up also with a $\tilde d = 1$. For the 10-dimensional strings with $\tilde d = 6$ to 4-dimensional black holes, we need to do one diagonal (double) dimensional reduction ending up with the same $\tilde d = 6$ and then 5 step by step direct dimensional reductions to give $\tilde d = 6 \to \tilde d - 5 = 1$, once again ending up with a $\tilde d = 1$.  In general, for the 10-dimensional p-branes with $0 \le p \le 6$  and $\tilde d = 7 - p$ to 4-dimensional black holes, we need to have p diagonal (double) dimensional reductions ending up with the same $\tilde d = 7 - p$ and then $6 - p$ step by step direct dimensional reductions to give $\tilde d = 7 - p \to \tilde d - (6 - p) = 1$, ending up with $\tilde d = 1$.   So this provides another way to understand why we end up with the same phase structure of $a^2 = 3$ charged black holes in 4-dimensions even though we begin with different phase structures in 10-dimensions.  In general, a direct 
dimensional reduction changes $\tilde d \to \tilde d - 1$ but may not change the underlying phase structure. For example, in the above consideration, if we consider to reduce the 10-dimensional 0-branes to 6-dimensional charged black holes instead,  then the phase structure remains characteristically unchanged since the resulting 
$\tilde d = 3$ at the end.    

Next, let us move to discuss the other set of systems, namely, the threshold $(D(p - 4), Dp)$ systems\footnote{The phase structures of the non-threshold systems $(D(p - 2), Dp)$ systems with $2 \le p \le 6$, considered in \cite{Lu:2012rm}, are the same as the simple constituent p-branes in the bound states. They are not relevant to the 4-dimensional charged black holes with the dilaton coupling  $a = 1$ which are threshold bound state of $a = \sqrt{3}$ charged black holes.} with $4\le p \le 6$ given in
 \cite{Lu:2012rm}.  We expect that these systems should correspond to the 4-dimensional black holes with $a = 1$.  From the appendix of \cite{Lu:2012rm}, we have the key quantity, the reduced inverse temperature function,  for the phase structures of these systems 
 \be
\label{gbf} b_{q_{p - 4}, q_p} (x) = \frac{x^{1/2}}{7 - p}
\left(\frac{\triangle_+}{\triangle_-}\right)^{1/2} \left(1 -
\frac{\triangle_+}{\triangle_-}\right)^{\frac{2 -\tilde d}{2\tilde d}}\left(1 + \frac{1 - G_{p - 4}^{ -
1}}{\frac{\triangle_-}{\triangle_+} - 1}\right)^{1/2}, 
\ee
where $\tilde d = 7 - p$, $q_p < x < 1$ and
\be\label{dpdm}  \triangle_+ =  1 - x, \qquad \triangle_- =  1 -
\frac{q^2_p}{x}.
\ee 
In the above,
\be
1 - G_{p - 4}^{- 1}  = 
\frac{1}{2}\left[\sqrt{\left(\frac{\triangle_-}{\triangle_+} -
1\right)^2 + 4  q_{p - 4}^2 \frac{\triangle_-}{\triangle_+}} -
\left(\frac{\triangle_-}{\triangle_+} - 1\right)\right].
\ee
The actual phase structure of the underlying system depends on whether  the function $b_{q_{p - 4}, q_p} (x)$ is divergent, non-vanishing finite or vanishing when $x \to q_p$, i.e., the lower end limit or the extremal limit. Let us examine this. When $x \to q_p$, $ \triangle_- /\triangle_+ \to 1$ and $1 -  G_{p - 4}^{- 1} \to 2 q_{p - 4} \neq 0$.
So we have 
\be 
 b_{q_{p - 4}, q_p} (x \to q_p) \sim \left(1 -
\frac{\triangle_+}{\triangle_-}\right)^{\frac{2 -\tilde d}{2\tilde d} - \frac{1}{2}},
\ee
where we give only the possible divergent factor.  So $b_{q_{p - 4}, q_p} (x \to q_p) $ diverges if $\tilde d > 1$, therefore having the van der Waals-Maxwell liquid-gas type phase structure.  For $\tilde d = 1$, $ b_{q_{p - 4}, q_p} (x \to q_p) $ is now non-vanishing finite, therefore having a phase structure of the simple charged black 5-branes. In other words, adding delocalized $(p - 4)$-branes to p-branes in the cases for $p = 5, 6$ changes their own phase structures, respectively.   Follow what we did for the simple charged black p-branes via diagonal (double) plus possible direct dimensional reductions, we end up 4-dimensional charged black holes all with $\tilde d = 1$. 
Among these, again only for $p = 6$, corresponding to $\tilde d = 1$, the reductions are all diagonal (double) ones and so the phase structure is the one of $a^2 = 1$ charged black holes in 4 dimensions.  Let us check this directly.  In diagonal (double) dimensional reductions, $\tilde d$ remains unchanged. So the behavior of $b_{q_2, q_6} (x \to q_6)$ remains the same as in 10 dimensions when reduced to 4 dimensions. In comparison with the corresponding 4-dimensional charged black hole behavior given in (\ref{bfunction}), we must have 
\be
\frac{a^2 - 1}{a^2 + 1} = \frac{2 - \tilde d}{2 \tilde d} - \frac{1}{2},
\ee
where $\tilde d = 1$.  This equation immediately gives $a^2 = 1$, as expected.  
 
   Before going further, let us make a few remarks: 1) Those charged black systems related by only diagonal (double) dimensional reductions have characteristically the same phase structure. 2)   The other reason for this is that for such related systems, they have the same $\beta^*$, the inverse temperature at infinity, such as that in 
(\ref{iT}) for 4-dimensional charged black holes. The  behavior of the reduced inverse temperature function, $b (x)$, whether divergent or not, is completely due to that of $\beta^*$. 3) For each given bulk dimension D, different brane systems have different phase structures as seen above. For examples, in 10-dimensions, the charged black p-branes have the van der Waals-Maxwell liquid-gas type phase structure for $0\le p \le 4$, the phase structure of the 4-dimensional charged black holes with $a^2 = 1$ for $p = 5$ and that of the 4-dimensional charged black holes with $a^2 = 3$ for $p = 6$.   We also see that the phase structure of 5-branes can be changed to have the van der Waals-Maxwell liquid-gas type if we add delocalized 1-branes to the 5-branes. We can also change that of 6-branes to have this phase structure by adding delocalized 0-branes to them \cite{Lu:2013nt, Lu:2014dra} even though the resulting system $(D0, D6)$ is not supersymmetric in the extremal limit. 
Actually this system gives rise to $a^2 = 0$ dyonic black holes in 4 dimensions \cite{Lu:2013nt}.  In other words, all possible phase structures in 10-dimensions for various well-defined different p-brane systems can be realized simply by different systems of 6-branes formed by 6-branes and other delocalized lower dimensional branes.  These systems have the same phase structures of 4-dimensional charged black holes with different dilaton coupling $a$. This further says that 4-dimensional 
charged black holes with different dilaton couplings capture all the possible phase structures of 10-dimensional charged brane systems under  conditions of similar settings. 

We now provide more evidence in support of the three points given above by considering the intersecting branes in 11 dimensions which are related to 4 dimensional charged black holes with $a^2 = 0, 1/3, 1, 3$ or $\Delta = 4/N = 1, 4/3, 2, 4$, respectively.  We know that there are many intersecting branes in 11 or 10 dimensions which can serve this purpose. Concretely, we will consider here the 2$\perp$2$\perp$5$\perp$5 system with 3 overall transverse dimensions whose extremal solution  was given first in \cite{Klebanov:1996mh} and the corresponding black one as well as the corresponding 4-charge 4-dimensional black holes was latter given in \cite{Cvetic:1996gq}.  The solution for the charged black  2$\perp$2$\perp$5$\perp$5 system in 11 dimensions from  \cite{Cvetic:1996gq} is, for the metric
\bea \label{11metric}
&&ds_{11}^2 = (T_1 T_2)^{-1/3} (G_1 G_2)^{-2/3}\left[- T_1 T_2 G_1 G_2 f d t^2 + G_1 \left(T_1 d y_1^2 + T_2 d y_3^2\right)\right. \nn
&&\qquad\quad \left. + G_2 \left(T_1 d y_2^2 + T_2 d y_4^2\right) + G_1 G_2 \left(dy_5^2 + dy_6^2 + dy_7^2\right) + f^{-1} d r^2 + r^2 d\Omega_2^2\right],
\eea
and for the 4-form field
\bea 
F_4 &=& - 3 dt \wedge\left(d T'_1 \wedge dy_1 \wedge dy_2 + dT'_2 \wedge dy_3\wedge dy_4\right) \nn
         &\,& + 3 \left(* dG'^{-1}_1 \wedge dy_2 \wedge dy_4 +  *dG'^{-1}_2 \wedge dy_1 \wedge dy_3\right).
\eea
In the above, one M2 brane is along $y_1, y_2$ directions and the other along $y_3, y_4$ directions, intersecting at a point while the two M5 branes intersect at a 3-brane along $y_5, y_6, y_7$. In addition, one M5 is along $y_1, y_3$ and the other along $y_2, y_4$ such that any M2 and any M5 intersect at a 1-brane.  They all obey the known intersecting rules. So we have isometries along the $1, 2, 3, 4, 5, 6, 7$ directions tangent to the branes, which can be used to reduced this system to 4-dimensional charged black holes via diagonal (double) dimensional reductions.  The above function $f$, parametrizing a derivation from the extremality, and functions $T_i, T'_i$ and $G_i, G'_i$, specifying the (non-extreme) M2 and M5 configuration, depend on the radial coordinate of  3-dimensional  overall transverse space,
\bea\label{ftg}
&&f = 1 - \frac{\mu}{r}, \qquad T_i^{-1} = 1 + \frac{{\cal Q}_i}{r}, \qquad T'_i = 1 - \frac{Q_i}{r} T_i,    \nn  
&&\quad {\cal Q}_i = \mu \sinh^2 \delta_i, \qquad Q_i = \mu \sinh\delta_i \cosh\delta_i, \qquad i = 1, 2, \\
&&\qquad G^{-1}_i = 1 + \frac{{\cal P}_i}{r}, \qquad G'^{-1}_i = 1 + \frac{P_i}{r},\nn
&&\quad {\cal P}_i = \mu \sinh^2\gamma_i, \qquad P_i = \mu \sinh\gamma_i \cosh\gamma_i, \quad i = 1, 2.
\eea
Upon diagonal (double) dimensional reductions to 4 dimensions, one finds the following Einstein-frame metric \cite{Cvetic:1996gq} as
\be \label{rd4dmetric}
ds_4^2 = - h(r) f(r) d t^2 + h^{-1} (r)\left[f^{-1}(r) d r^2 + r^2 d\Omega_2^2\right],
\ee
where 
\be 
h(r) = (T_1 T_2 G_1 G_2)^{1/2} = \frac{r^2}{\left[(r + {\cal Q}_1) (r + {\cal Q}_2)(r + {\cal P}_1)(r + {\cal P}_2)\right]^{1/2}}.
\ee
Note that similar solutions for 1-charge, 2-charge and 3-charge cases can be obtained from the above by setting vanish 3 charges, 2 charges and 1 charge from the above, respectively.  To be consistent with the 4-dimensional charged black holes with different dilaton coupling given in (\ref{bhsolution}), we need to set the remaining 
non-vanishing charges equal, denoted as ${\cal Q}$, respectively\footnote{The same conclusion can also be reached even if we leave the non-vanishing charges arbitrary.}.  Let us first look at the black holes given in (\ref{rd4dmetric}).  We find that if sending $r \to r - {\cal Q}$ in the metric with the ${\cal Q}$ just defined, we have the physical radius of the 2-sphere as
\be \label{pr}
R (r) = r \left(1 - \frac{\cal Q}{r}\right)^{1 - \frac{N}{4}},
\ee
where $N$ denotes the number of non-vanishing charges before we set them equal.  Comparing this with the one given in (\ref{bhone}), we have immediately the expected relation,
\be \label{dcone}
a^2 = \frac{4}{N} - 1,
\ee 
if we identify $r_- = {\cal Q}$. Let us further check the consistency for the metric with the dilaton coupling and the identification of $r_-$.  We have now
\be
h (r) =    \left(1 - \frac{r_-}{r}\right)^{N/2}, \qquad f(r) = \left(1 - \frac{r_-}{r}\right)^{- 1}  \left(1 - \frac{r_+}{r}\right)
\ee
where the expected $r_+ = {\cal Q} + \mu$.  Then we have
\be 
h(r) f(r) = \left(1 - \frac{r_+}{r}\right)  \left(1 - \frac{r_-}{r}\right)^{N/2 - 1} ,
\ee
which agrees with the $\lambda^2$ given in (\ref{bhone}) when the dilaton coupling satisfies (\ref{dcone}).  With all these, the metric (\ref{rd4dmetric}) agrees completely with that given in (\ref{bhsolution}).  Then the inverse temperature given in (\ref{iT}) and the local inverse temperature given in (\ref{lT}) both hold here, too.  

   Now let us look at the 11-dimensional metric (\ref{11metric}) with the same conditions as given for the 4-dimensional black holes just described.  Consider the metric in Euclidean signature and require the metric free of a conical singularity at $r = r_+$, we end up with the corresponding Euclidean time with a period 
\be 
\beta^*_{11} = 4 \pi r_+ \left(1 - \frac{r_-}{r_+}\right)^{\frac{2 - N}{2}},
\ee
where we add an subscript $11$ for the inverse temperature of the 11-dimensional case.   This $\beta^*_{11}$ agrees precisely with the one for the 4-dimensional charged black holes given in (\ref{iT}) once (\ref{dcone}) holds.  Therefore the 11-dimensional charged black intersecting branes have characteristically the same phase structure as the corresponding 4-dimensional charged black holes related via diagonal (double) dimensional reductions.   

In summary, a diagonal (double) dimensional reduction preserves the underlying phase structure in general while a direct dimensional reduction has the potential to change the underlying phase structure, which depends on the value of $\Delta$ and the small enough resulting $\tilde d$.   This very fact implies that a higher dimensional charged black system has characteristically the same phase structure as the corresponding charged black hole if former is related to the latter purely by 
diagonal (double) dimensional reductions. As discussed earlier in point 3, this further implies that the 4-dimensional charged black holes with all possible dilaton coupling $a^2 \ge 0$ can capture all possible phase structures of higher dimensional systems, simple or complicated/known or unknown, under conditions of similar settings. In other words, the phase structures of these higher dimensional systems are just those discussed in this paper for the 4-dimensional charged black holes with an arbitrary dilaton coupling.

\section{Summary and Conclusion}
In this paper we study the thermodynamics and the phase structures of the 4-dimensional asymptotically flat dilatonic black holes, placed in a cavity {\it a la} York, in string theory. We considered these charged black systems in canonical ensemble  in which the temperature as well as the dilaton, at the wall of cavity, and the amount of charge inside the cavity are fixed. We employed the Euclidean action formalism to compute the Helmholtz free energy and to analyze the thermal stability of the underlying thermodynamical system. We find that the dilaton coupling plays a key role in determining the underlying phase structure of the charged black system. We also make connections of the phase structures of these systems to higher dimensional charged black p-branes via diagonal (double) and/or direct dimensional reductions. 

For the chargeless case, similar to higher dimensional black branes, the dilaton coupling is irrelevant to the underlying phase structure. This phase structure has been discussed in section 4.1 and  is characteristically the same as that of Schwarzschild black holes .   

The focus of this paper is on the 4-dimensional charged black holes with an arbitrary dilaton coupling $a$. This coupling is determined by the value of $\Delta$, characterizing different kinds of brane systems in higher dimensions. In general, we find three kinds of phase structure, depending on $a < 1, a = 1$ or $a > 1$.  The $a < 1$ case gives rise to the van der Waals-Maxwell liquid-gas type phase structure.  The phase structure of the $a > 1$ case resembles that of charged black 6-branes studied previously and it is discussed in section 4.2. The $a = 1$ case gives a phase structure which is characteristically the same as that of charged black 5-branes studied previously, being viewed as a borderline in phase structure between the $a < 1$ and $a > 1$ cases, and it is discussed in detail in section 4.2.   In other words, the dilaton coupling $a$ decides the underlying phase structure. 

The 4-dimensional black holes with an arbitrary dilaton coupling are connected to the higher dimensional brane systems, simple or intersecting/known or unknown, via diagonal (double) and/or dimensional reductions. In particular, we find that the phase structure is preserved under diagonal (double) dimensional reductions while this may not hold true under direct dimensional reductions.  Since the dilaton coupling is left as an arbitrary parameter in our study of the underlying phase structures for the 4-dimensional charged black holes, we provide evidence supporting that the uncovered phase structures given in this paper have the potential to describe all possible phase structures, under conditions of similar settings, of the higher dimensional well-defined brane systems, some of which may be difficult to have explicit solutions or difficult to analyze or not even be known up to now.   This is the advantage and usefulness of the present study.

\section*{Acknowledgements}
We would like to thank Hong Lu for fruitful discussions.  We acknowledge support by a key grant from the NSF of China
with Grant No: 11235010.


\bibliographystyle{unsrt}
\bibliography{QiangJia}
\end{document}